%% file: compass_l1551irs5_cd3cn.tex
\documentclass{aa}  

\usepackage{graphicx}
\usepackage{txfonts}
\usepackage{lipsum}
\usepackage{subcaption}         
\usepackage{lscape}             
\usepackage{placeins}           
                                
\bibpunct[]{(}{)}{;}{a}{}{,}

\begin{document}

   \title{Complex Organic Molecules in Protostars\\ with ALMA Spectral
Surveys (COMPASS)}

   \subtitle{VIII. First interstellar detection of fully deuterated methyl cyanide}
   \titlerunning{COMPASS: First interstellar detection of fully deuterated methyl cyanide}

%
%
%

   \author{A. Belloche\inst{1}\corrauth{belloche@mpifr-bonn.mpg.de}
        \and J.~K. J{\o}rgensen\inst{2}
        \and M.~N. Drozdovskaya\inst{3}
        \and S. Spezzano\inst{4}
        \and A. Coutens\inst{5}
        \and J.-E. Lee\inst{6}
        \and\\ F. Cruz-S{\'a}enz de Miera\inst{5}
        \and D. Harsono\inst{7}
        \and J.-H. Jeong\inst{6}
        \and S. Maret\inst{8}
        \and B.~A. McGuire\inst{9,10}
        \and\\ P. Nazari\inst{11,12}
        \and A.~L. Plunkett\inst{10}
        \and Y.-L. Yang\inst{13}
        \and H.-S. Yun\inst{6}
        }

   \institute{Max-Planck-Institut f{\"u}r Radioastronomie, 
              Auf dem H{\"u}gel 69, 53121 Bonn, Germany
            \and
              Niels Bohr Institute, University of Copenhagen, Jagtvej 155A, DK 
              2200 Copenhagen N, Denmark
            \and 
              Department of Chemistry, Biochemistry and Pharmaceutical
              Sciences, Universit{\"a}t Bern, Freiestrasse 3, 3012 Bern, 
              Switzerland
            \and
              Max-Planck-Institut f{\"u}r extraterrestrische Physik, 
              Giessenbachstrasse 1, 85748 Garching bei M{\"u}nchen, Germany 
            \and
              Univ. Toulouse, CNES, CNRS, IRAP, Toulouse, France
            \and
              Department of Physics and Astronomy, Seoul National University, 
              1 Gwanak-ro, Gwanak-gu, Seoul 08826, Republic of Korea
            \and
              Institute of Astronomy, Department of Physics, National Tsing Hua 
              University, No. 101, Section 2, Kuang-Fu Road, Hsinchu 30013, 
              Taiwan, R.O.C.
            \and
              Univ. Grenoble Alpes, CNRS, IPAG, 38000 Grenoble, France
            \and
              Department of Chemistry, Massachusetts Institute of Technology, 
              77 Massachusetts Ave, Cambridge, MA 02139 USA
            \and 
              National Radio Astronomy Observatory, 520 Edgemont Road, 
              Charlottesville, VA 22903, USA
            \and
              Department of Physics, University of Warwick, CV4 7AL, Coventry,
              UK
            \and
              Institute of Astronomy, University of Cambridge, Madingley Rd, 
              CB3 0HA, Cambridge, UK
            \and
              Star and Planet Formation Laboratory, RIKEN Pioneering Research 
              Institute, Wako-shi, Saitama, 351-0198, Japan
        }

   \date{Received 12 August 2026 / Accepted 28 October 2026}

 
  \abstract
   {Deuteration is a powerful metric to gain insight into the formation pathways
of interstellar molecules in star forming regions.}
   {Our primary goal is to search for the fully deuterated form of methyl 
cyanide, CD$_3$CN, in the interstellar medium (ISM).} 
   {We used a sensitive imaging spectral survey of the protostellar system 
L1551 IRS5 in the Taurus molecular cloud performed as part of the Large 
Program carried out with the Atacama Large Millimeter/submillimeter Array 
(ALMA) called Complex Organic Molecules in Protostars with ALMA Spectral 
Surveys (COMPASS). We derived the column densities of methyl cyanide and 
several of its less abundant isotopologs, including deuterated ones, under the 
assumption of local thermodynamic equilibrium.}
   {All three deuterated isotopologs of methyl cyanide were detected toward 
the northern binary component of L1551~IRS5. In particular, we report the first 
interstellar detection of CD$_3$CN. The measured D/H ratio increases by about 
half an order of magnitude from one deuterated isotopolog to the next, in a 
similar way as what was measured for methanol in several other low-mass 
protostars. This suggests that methyl cyanide predominantly forms like 
methanol on the surface of dust grains during the cold prestellar phase. 
However, the measured factor of 2 difference between the D/H ratios of methyl 
cyanide and methanol may either indicate that gas-phase processes during the 
warm-up stage contribute approximately equally to the production of methyl 
cyanide, or that methyl cyanide and its deuterated isotopologs are formed 
during the prestellar phase at a somewhat earlier time than methanol.}
   {Multiply deuterated isotopologs are crucial to understand the formation and 
inheritance of molecules in the ISM by coupling observations and astrochemical 
models. Unfortunately, no model has so far investigated the deuteration of 
methyl cyanide up to the fully deuterated form. More advanced models will be 
needed to test whether methyl cyanide is predominantly formed and deuterated on 
the surface of dust grains and inherited from the prestellar phase, as the
observational results may suggest.}


   \maketitle

\section{Introduction}
\label{s:introduction}

In the field of astrochemistry, complex organic molecules (COMs) are defined as
carbon-bearing species made of at least six atoms \citep[][]{Herbst09}.
Understanding how and when these molecules form in the interstellar medium 
(ISM) and if the chemical composition of the material involved in star 
formation is inherited or altered from one evolutionary phase to the next
are some of the main goals of astrochemistry
\citep[see, e.g.,][]{Jorgensen20}. One property of COMs that can help us 
decipher their origin is their degree of deuteration. Molecules get enriched 
in deuterium during the cold and dense prestellar phase thanks to the 
exothermicity of the reaction of the main molecular ion, H$_3^+$, with the 
molecular reservoir of deuterium, HD \citep[e.g.,][]{Ceccarelli14}. 
Measuring the deuterium fractionation of a molecule (that is the column 
density ratio of its deuterated form to its hydrogenated form) can thus tell 
us about the time when, and conditions (e.g., temperature) under which, a 
molecule has formed.

In this work, we are interested in the deuteration of methyl cyanide, CH$_3$CN.
This species is one of the simplest COMs in the family of nitrogen-bearing 
molecules. Both solid-phase and gas-phase processes have been proposed for the 
formation of methyl cyanide. For instance, in the astrochemical models of 
\citet{Aikawa12} and \citet{Taquet14}, it predominantly forms on the surface 
of dust grains, both during the cold, prestellar phase and later during the 
warm-up phase once the protostar has started to heat up its envelope. The 
tentative identification of methyl cyanide in interstellar ices gives some 
support to these model predictions \citep[][]{Nazari24}. In contrast, in the 
astrochemical model of \citet{Garrod22}, gas-phase reactions significantly 
contribute to the formation of methyl cyanide during the warm-up phase, during 
which deuteration is expected to be inefficient. Measuring the deuteration 
level of methyl cyanide in protostellar envelopes can thus allow us to 
discriminate the reaction schemes proposed by these models and better 
understand the formation of this molecule. So far, two of its three deuterated 
isotopologs have been detected in the ISM: CH$_2$DCN and CHD$_2$CN 
\citep[e.g.,][]{Gerin92,Calcutt18}.

The Large Program performed with the Atacama Large Millimeter/submillimeter 
Array (ALMA) called Complex Organic Molecules in Protostars with ALMA Spectral 
Surveys \citep[\hbox{COMPASS};][]{Jorgensen26} is an unbiased spectral survey 
of 11 low-mass Class 0 and I protostellar sources that are known to be 
chemically rich and reside in diverse environments. The goal of the COMPASS 
project is to characterize the chemical composition of these sources and 
investigate the impact of the environment and physical evolution of a star 
forming core on its chemical evolution. One of the targeted sources is the 
Class~I protostellar system L1551~IRS5 \citep[see][ for an overview of its 
properties]{Jorgensen26}. This binary system is located in the Taurus
molecular cloud at a distance of 145~pc \citep[][]{Zucker20,Jorgensen26}. The
binary components, called L1551~IRS5N and IRS5S \citep[][]{Lim16}, are 
separated by about 50~au and each one possesses a compact (radius of about 
10~au) circumstellar disk \citep[][]{Rodriguez98}. The system
is surrounded by a circumbinary disk or flattened inner envelope with a radius 
of $\sim$150~au \citep[][]{CruzSaenzDeMiera19,Takakuwa20} and a larger scale 
(radius $\sim$0.1~pc)
$\sim$1--2~$M_\odot$ envelope \citep[][]{Jorgensen02,Chou14}. L1551~IRS5 is 
known to have a relatively high D/H ratio of water compared to other protostars
\citep[][]{Andreu23}. 
From a preliminary analysis of the methyl cyanide emission in the 
\hbox{COMPASS} sources, L1551~IRS5 stood out as a particularly interesting 
source with respect to the deuteration of this molecule. This motivated a 
detailed study of methyl cyanide and its isotopologs toward L1551~IRS5, which 
is presented in this article.

The article is structured as follows. The observational setup and the method 
used to extract column densities from the spectra are presented in 
Sect.~\ref{s:observations}. We report the results in Sect.~\ref{s:results}
and discuss them in Sect.~\ref{s:discussion}. The conclusions are stated in 
Sect.~\ref{s:conclusions}.

\section{Observations and radiative transfer modeling}
\label{s:observations}

\subsection{ALMA observations}
\label{ss:alma}

We used data collected between 2023 December and 2024 July with the 12~m array 
of ALMA toward L1551~IRS5 as part of the COMPASS Large Program
\citep[][]{Jorgensen26}\footnote{https://erda.ku.dk/vgrid/COMPASS/}. The 
COMPASS data were reduced with a dedicated pipeline archived on Zenodo 
\citep{Maret25}. We refer the reader to \citet{Plunkett26} for details about
the observations, the data reduction strategy, and the subtraction of the 
continuum emission. L1551~IRS5 was observed with nine setups named S1 to S9 
that fully cover the frequency range from 279.0 to 311.7~GHz with an effective
spectral resolution of 488~kHz (0.52--0.47~km~s$^{-1}$). The channel spacing 
was 244~kHz. Each setup consists of four spectral windows labeled W25, W27, W29,
and W31 that slightly overlap each other. The beam size (half-power beam 
width, HPBW) varies between $0.15\arcsec \times 0.11\arcsec$ and 
$0.20\arcsec \times 0.13\arcsec$, with median values of 
$0.16\arcsec \times 0.13\arcsec$ (23~au $\times$ 19~au). L1551~IRS5 is one of 
the two COMPASS sources that were on purpose observed at higher angular 
resolution because of their tight binarity. The rms noise level per channel 
ranges from 1.5~mJy~beam$^{-1}$ (S1W29, S8W25) to 3.6~mJy~beam$^{-1}$ (S7W27), 
with a median value of 2.0~mJy~beam$^{-1}$.

\subsection{Radiative transfer modeling}
\label{ss:weeds}

We followed the same strategy as \citet{Belloche26} to analyze the
spectra under the assumption of local thermodynamic equilibrium 
\citep[LTE; see][ for a justification of this approximation for the 
COMPASS sources]{Coutens26}. The 
modeling was performed with Weeds \citep[][]{Maret11}, an extension of the 
CLASS software that is part of 
GILDAS\footnote{https://www.iram.fr/IRAMFR/GILDAS}. Weeds takes the line 
opacities, finite angular resolution of the data, and continuum emission into 
account to compute the synthetic spectra. 
The rotational spectrum of each molecule was modeled with five parameters: the 
emission size, $\theta_{\mathrm s}$, column density, $N$, rotational 
temperature, $T$, local-standard-of-rest (LSR) velocity\footnote{The actual 
parameter used in Weeds is the velocity offset with respect to the LSR velocity 
stored in the header of the spectra.}, $V$, and line width (full width at half 
maximum, FWHM), $\Delta V$.

\subsection{Spectroscopy}
\label{ss:spectro}

The spectra of methyl cyanide and its isotopologs were modeled using 
spectroscopic entries taken from  the Cologne Database for Molecular 
Spectroscopy 
\citep[CDMS\footnote{https://cdms.astro.uni-koeln.de},][]{Mueller05}. These
entries are mainly based on \citet{Mueller15} for CH$_3$CN, \citet{Mueller26}
for $^{13}$CH$_3$CN and CH$_3$$^{13}$CN, \citet{Mueller09} and 
\citet{Mueller16} for CH$_3$C$^{15}$N, \citet{Nguyen13} for CH$_2$DCN and 
CHD$_2$CN, and \citet{Demaison79} and \citet{Brown89} for CD$_3$CN. The 
partition function of the CDMS entry of CH$_3$CN already accounts for the
vibrational contribution. This is not the case for the other isotopologs.
We used the information provided in the CDMS to compute the vibrational
contribution to the partition function and correct their column densities 
accordingly, except for CHD$_2$CN and CD$_3$CN. For the latter isotopologs,
we used the vibrational energies of CD$_3$CN published in 
Table 1 of \citet{Duncan78} to compute the correction with
Equation 1 of \citet{Belloche25}. The list of energies (in cm$^{-1}$) is: 
334.80, 334.80, 831.33, 847.11, 847.11, 1046.45, 1046.45, 1110.00, 2127.80, 
2256.56, 2256.56, and 2277.88. Identical values correspond to doubly degenerate
states, as indicated in the CDMS documentation on CH$_3$CN.

\section{Results}
\label{s:results}

\subsection{Source selection}
\label{ss:source}

\begin{figure}[!t]
\centerline{\resizebox{0.90\hsize}{!}{\includegraphics[angle=0]{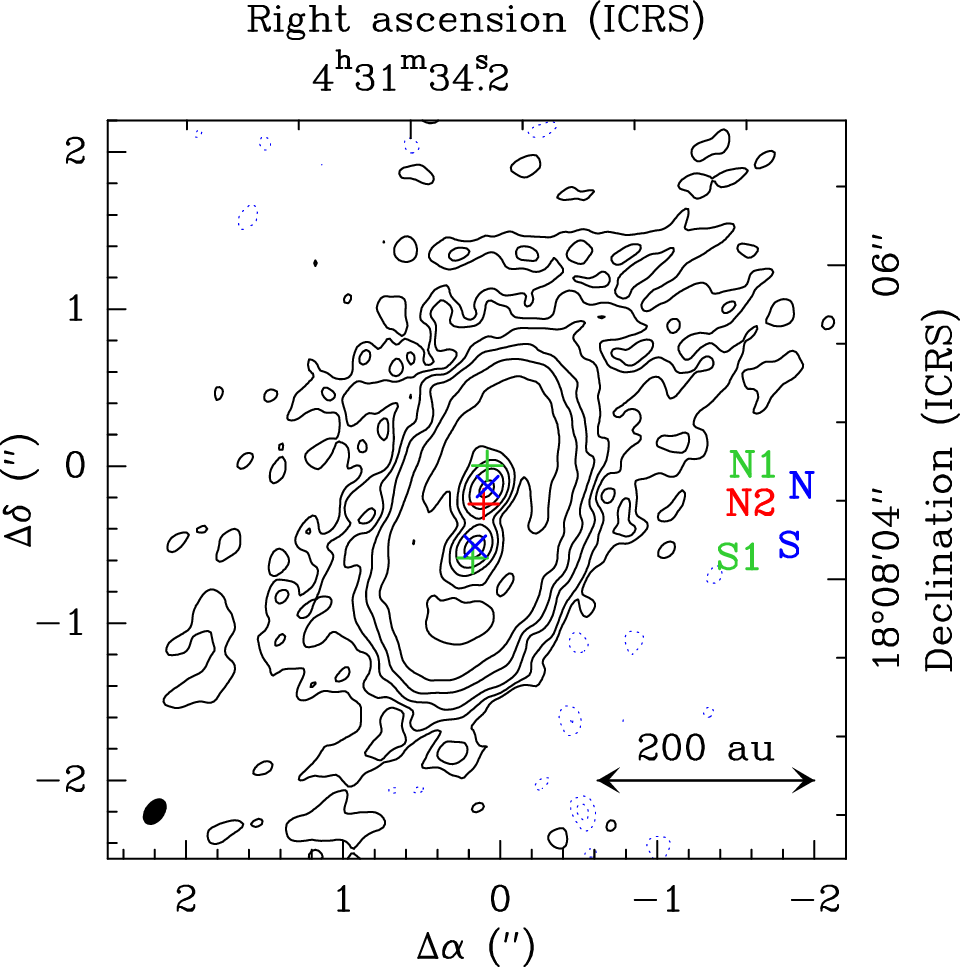}}}
\caption{Continuum emission map at $\sim$279.5~GHz (S1W25). The black contours 
start at 3$\sigma$ and then increase by a factor of two, with $\sigma$ the
noise level (0.12 mJy~beam$^{-1}$). The blue dotted contours show the 
$-3\sigma$ and $-6\sigma$ levels. The blue crosses indicate the continuum 
peak positions of the binary components, L1551~IRS5N and L1551~IRS5S. The plus 
symbols mark the three channel count peak positions reported in Table 3 of
\citet{Jorgensen26}: L1551~IRS5N2 in red and L1551~IRS5N1 and L1551~IRS5S1 in 
green. The beam is displayed in the bottom left corner.}
\label{f:contmap}
\end{figure}

To identify reference positions for the analysis of the COM emission, the 
COMPASS team created maps of channel counts for all COMPASS sources 
\citep[see][]{Jorgensen26}. These maps show, for each pixel, the number of 
channels of the continuum-subtracted spectra that have a flux density higher 
than 6 times the rms noise level across the full spectral survey, excluding 
channels that contain emission from species with strong and extended emission. 
These maps reveal the locations of hot corino emission in the COMPASS fields. 
In the case of L1551~IRS5, three peaks of channel counts were identified (see 
Fig.~\ref{f:contmap}). Two of them, L1551~IRS5N1 and L1551~IRS5N2, are located 
0.13$\arcsec$ to the north and 0.11$\arcsec$ roughly to the south of the main 
continuum peak, L1551~IRS5N, respectively. The third channel count peak, 
L1551~IRS5S1, is offset by 0.08$\arcsec$ from the secondary continuum peak, 
L1551~IRS5S. The offsets between the continuum and channel count peaks 
are likely due to high opacities of the continuum emission obscuring the line 
emission \citep[][]{Takakuwa20,Kang26}. In this work, we analyzed the 
spectrum toward L1551~IRS5N2, which is located at the ICRS equatorial position
$04^{\rm h}31^{\rm m}34\fs1674$, $18^\circ08\arcmin04\farcs48$ (red cross in 
Fig.~\ref{f:contmap}). In practice, we used the spectrum of the pixel
closest to the channel count peak. No spatial averaging was performed.

\subsection{Derivation of LTE parameters}
\label{ss:param}

Our best-fit synthetic spectra of CH$_3$CN $\varv=0$, 
$\varv_8=1$\footnote{$\varv_8=1$ is the lowest vibrationally excited state of 
methyl cyanide.}, $\varv_8=2$, 
$^{13}$CH$_3$CN $\varv=0$, $\varv_8=1$, CH$_3$$^{13}$CN $\varv=0$, $\varv_8=1$,
CH$_3$C$^{15}$N $\varv=0$, $\varv_8=1$, CH$_2$DCN $\varv=0$, and CHD$_2$CN 
$\varv=0$ are displayed in 
Figs.~\ref{f:spec_ch3cn_ve0}--\ref{f:spec_chd2cn_ve0}, respectively, along 
with our preliminary LTE model that includes the contribution of all species 
that we have identified toward L1551~IRS5N2 so far, overlaid on the COMPASS 
spectrum observed toward L1551~IRS5N2.
Our preliminary model currently contains 34 molecules, 23 of their
deuterated isotopologs, 28 other isotopologs, and two dozens of
vibrational states of some of these molecules.

The best-fit model of methyl cyanide and its isotopologs was obtained 
following the same method as described in \citet{Belloche26} for the 
emission of methanol and its isotopologs toward the COMPASS source IRAS4A2. 
The systemic velocity and line width can be directly derived from Gaussian fits 
of uncontaminated, optically thin lines identified by visual inspection on the 
basis of our LTE model that provides line optical depths and degree of 
contamination by other species). We used a velocity of 5.8~km~s$^{-1}$
and a line width of 1.5~km~s$^{-1}$ for all methyl cyanide isotopologs. For
each isotopolog, we constructed a population diagram by only selecting lines 
that have an optical depth lower than 2 (on the basis of the best-fit model) 
and are not much contaminated by emission from other species 
(Figs.~\ref{f:popdiag_ch3cn}--\ref{f:popdiag_cd3cn}). The intensities were
computed by direct integration over the profile of the lines. The rotational 
temperatures derived from fits to these population diagrams, after correction 
for opacity and contamination\footnote{For a given transition 
of a given molecule, the contamination was computed as the difference between 
the integrated intensity of the synthetic spectrum that contains the 
contribution of all identified species and the integrated intensity of the 
synthetic spectrum of that molecule only. This number was then subtracted from
the integrated intensity of the observed spectrum.} on the basis of our 
best-fit LTE 
model, are listed in Table~\ref{t:popfit}. Within the uncertainties, the fit 
results point to a rotational temperature of $\sim$200~K, which is the value 
that we used for the LTE modeling. The only isotopolog that seems to deviate 
from this temperature is CHD$_2$CN with a fitted value of $\sim$130~K. 
However, the dispersion of the data points in the energy range of 120--180~K in 
Fig.~\ref{f:popdiag_chd2cn} may indicate that the upper data points in this
figure, which bias the fit toward lower temperature values, are contaminated 
by still unidentified species (transitions with $E_{\rm up}$ in this range of
energy are marked with a magenta cross in Fig.~\ref{f:spec_chd2cn_ve0}). A fit
restricted to the transitions with $E_{\rm up}/k$ higher than 180 K yields a
temperature of $213 \pm 45$~K. Therefore, we also assumed a temperature of 
200~K to model the emission of CHD$_2$CN.

\input{tab_l1551irs5_p2_cd3cn_popfit.tex}

The emission size of the methyl cyanide isotopologs is more difficult to 
determine. \citet{CruzSaenzDeMiera25} analyzed ALMA data obtained toward 
L1551~IRS5 at an angular resolution of $0.20\arcsec \times 0.19\arcsec$. They 
showed that methanol traces a clear velocity gradient from south to north 
around L1551~IRS5N. They performed an automatic fit of the COM emission toward
a position offset by $\sim$1$\arcsec$ to the northwest from the continuum 
peak of L1551~IRS5N in their data set. The fit was done with a Markov 
chain Monte Carlo method using both optically 
thick and thin lines of a dozen of COMs. The derived emission sizes range from 
0.05$\arcsec$ to 0.42$\arcsec$ depending on the COM, with a median size of 
0.08$\arcsec$. They obtained for instance sizes of 0.06$\arcsec$ and 
0.43$\arcsec$ for $^{13}$CH$_3$CN and CH$_3$C$^{15}$N, respectively. Such a 
large discrepancy between two isotopologs that in reality probably trace the 
same region indicates that these size measurements are likely more uncertain 
than suggested by the reported errors ($\sim$15--25$\%$). From a non-LTE 
analysis of another ALMA data set that targeted L1551~IRS5 at an angular 
resolution of $0.37\arcsec \times 0.31\arcsec$, \citet{Bianchi20} derived an 
emission size of 0.15$\arcsec$ for methanol and its $^{13}$C isotopolog toward
the continuum peak of L1551~IRS5N. 

Integrated intensity maps obtained from the COMPASS data for uncontaminated 
(at the position of L1551~IRS5N2)
transitions of the methyl cyanide isotopologs are displayed in 
Figs.~\ref{f:map_ch3cn}--\ref{f:map_cd3cn}. The spectra were integrated over a 
velocity range of 2.5--3~km~s$^{-1}$ centered on the systemic 
velocity of L1551~IRS5N2, which is blueshifted by about 2.8~km~s$^{-1}$ with 
respect to L1551~IRS5N 
\citep[$V_{\mathrm{sys}} \sim 8.6$~km~s$^{-1}$,][]{Marchand24}. Because of the 
pronounced south-north velocity gradient around L1551~IRS5N which is such that 
the velocity difference between L1551~IRS5N1 and L1551~IRS5N2 
($\sim$4~km~s$^{-1}$) is much larger than the line width, any emission that 
shows up to the north of L1551~IRS5N in 
Figs.~\ref{f:map_ch3cn}--\ref{f:map_cd3cn} must trace other molecules. This 
makes it more difficult to estimate the emission size of methyl cyanide and 
its isotopologs by fitting the spatial extent of their integrated emission. 
The deconvolved sizes that we derived from two-dimensional Gaussian fits to 
the integrated intensity maps of all isotopologs range from 0.10$\arcsec$ to 
0.51$\arcsec$. The smallest values are likely the ones that are the least 
affected by contamination, which would suggest an emission size on the order 
of 0.1$\arcsec$. Adopting a rotational temperature of 200~K as derived above
from the population diagrams, we need an emission size of 0.09$\arcsec$ to 
reproduce the peak intensity of the optically thick lines of CH$_3$CN. By 
fixing the emission size to this value for all isotopologs, the LTE modeling 
of the COMPASS spectra yields the parameters listed in the top part of 
Table~\ref{t:coldens}. Smaller sizes would lead to an underestimate of the
strength of the optically thick lines, which is not acceptable under the 
assumption that all transitions of the molecule trace the same emission 
size\footnote{Optically thick lines have a peak intensity that scales with the 
temperature and the beam filling factor. An underestimate of their peak 
intensity thus directly points to an underestimate of either the temperature 
or the emission size.}. 

\input{tab_l1551irs5_p2_cd3cn_weedsmodel.tex}

\begin{figure*}[!h]
\centerline{\resizebox{0.92\hsize}{!}{\includegraphics[angle=0]{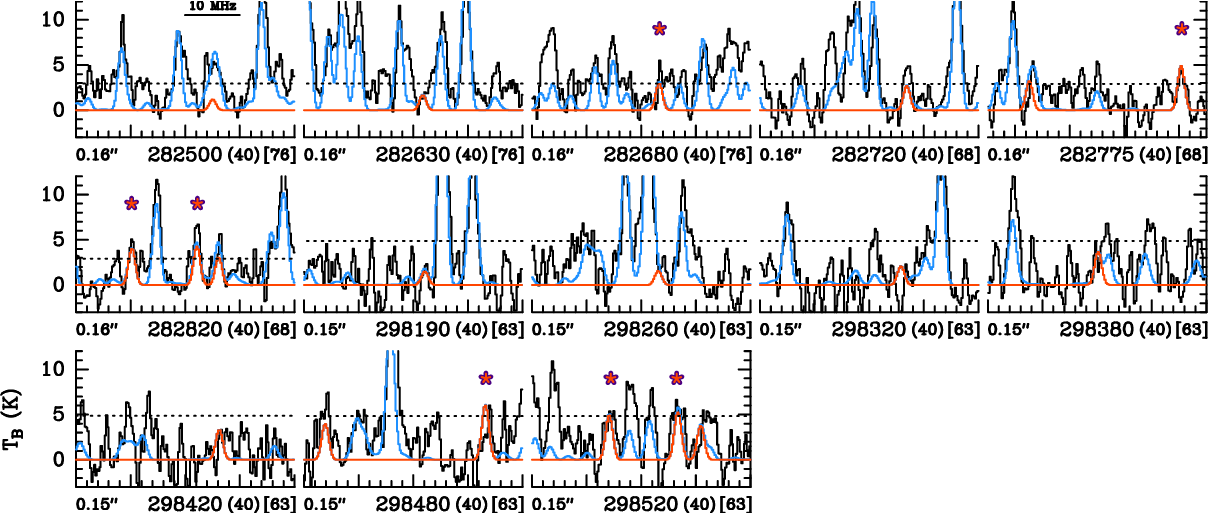}}}
\caption{Transitions of CD$_3$CN $\varv=0$ covered by the COMPASS survey
toward L1551~IRS5N2 with ALMA. The best-fit LTE synthetic spectrum of 
CD$_3$CN $\varv=0$ is shown in red and overlaid on the observed 
spectrum displayed in black. The blue synthetic spectrum contains the 
contribution of all molecules included in our model of L1551~IRS5N2 so far, 
including the molecule shown in red. The seven spectral lines of CD$_3$CN that 
have a peak signal-to-noise (S/N) ratio higher than 3 and are not 
significantly contaminated 
by emission from other (identified or unidentified) molecules, thereby firmly 
contributing to the identification of CD$_3$CN, are marked with red stars. 
The central frequency is indicated in MHz below each panel as well as the 
HPBW (geometrical mean) on the left, the width of each panel in MHz in 
parentheses, and the continuum level in K of the baseline-subtracted spectra 
in brackets. For guidance, the frequency interval between two big
tickmarks is indicated in the top left panel.
The y-axis is labeled in brightness temperature units (K). The 
frequency axis refers to the LSR velocity stored in the header of the spectra 
(9.1~km~s$^{-1}$). The dotted line indicates the 3$\sigma$ noise level.}
\label{f:spec_cd3cn_ve0}
\end{figure*}

Because the emission size is poorly constrained, we also computed two 
alternative models with an emission size of 0.15$\arcsec$ that corresponds to 
the size derived by \citet{Bianchi20}. For the first one, we kept the 
rotational temperature at 200~K and adjusted the column densities to fit the 
observed spectra. This model overestimates the strength of the optically 
thick lines of methyl cyanide (see Fig.~\ref{f:spec_ch3cn_ve0_alt1}). Our LTE 
model assumes uniform properties along 
the line of sight. In reality, we expect the excitation temperature to 
decrease toward the observer, meaning that optically thick lines are likely to
trace layers with lower excitation temperatures and thus be weaker than a 
uniform model would predict. Therefore, an overestimate of the strength of 
optically thick lines within the simple modeling framework (constant 
temperature along the line of sight) is not a sufficient criterion to reject a 
model. For the second alternative model (again with an emission size of 
0.15$\arcsec$), we reduced the rotational temperature to 150~K in order to
match the strength of the optically thick lines. Although the overall match of 
this second alternative model to the observations is somewhat better than the 
first alternative with respect to the optically thick lines, we do not 
think that it is a better model because its 
rotational temperature is not consistent with the temperature derived from the 
population diagrams (Table~\ref{t:popfit}). The derived parameters of both 
alternative models are listed in the middle and bottom parts of 
Table~\ref{t:coldens}. 

\subsection{Column densities of methyl cyanide isotopologs and first interstellar detection of CD$_3$CN}
\label{ss:coldens}

Figures~\ref{f:spec_ch3cn_ve0}--\ref{f:spec_chd2cn_ve0} show that rotational 
transitions of all the methyl cyanide isotopologs listed in 
Sect.~\ref{ss:param} are clearly detected, at least in the vibrational ground 
state. Lines from within $\varv_8=1$ are also detected for the main isotopolog 
(Fig.~\ref{f:spec_ch3cn_v8e1}). Lines of the main isotopolog from within 
$\varv_8=2$ (Fig.~\ref{f:spec_ch3cn_v8e2}) and lines of the $^{13}$C and 
$^{15}$N isotopologs from within $\varv_8=1$ 
(Figs.~\ref{f:spec_ch3cn_13c1_v8e1}, \ref{f:spec_ch3cn_13c2_v8e1}, and
\ref{f:spec_ch3cn_15n_v8e1}) contribute to the observed signal but they are too 
weak to allow us to identify these vibrational states unambiguously. The column
densities derived from our best-fit model with an emission size of 
0.09$\arcsec$ and the two alternative models with a size of 0.15$\arcsec$
are listed in Table~\ref{t:coldens}. This table also provides the number of 
detected lines for each isotopolog. These correspond to lines that have a peak
temperature higher than 3 times the noise level, coincide in frequency with a 
clear peak in the observed spectrum, and are not severely blended with 
emission from other molecules. In practice, the latter criterion means that we
counted a line as detected if its modeled peak temperature represents at least
two thirds of the detected signal, the rest being either filled by emission
of a contaminating molecule included in our full model, or still unidentified.

We not only detected the singly and doubly deuterated isotopologs of methyl 
cyanide but also the fully deuterated one,  CD$_3$CN, with seven lines 
detected above the 3$\sigma$ level as marked with red stars in 
Fig.~\ref{f:spec_cd3cn_ve0}. This is the first interstellar detection of fully 
deuterated methyl cyanide. The small (1.5$\sigma$) discrepancy between 
synthetic and observed spectra in the second to last panel of 
Fig.~\ref{f:spec_cd3cn_ve0} is likely a combination of noise and a slight 
overestimation of the baseline level. Because of the limited signal-to-noise 
ratio of the detected lines, the population diagram of CD$_3$CN is poorly 
constrained (see Fig.~\ref{f:popdiag_cd3cn} and Table~\ref{t:popfit}) but a 
temperature of 200~K, as derived from the other isotopologs, is consistent 
with the observed spectrum. The integrated intensity maps of a selection of 
four of these transitions show compact emission peaking on L1551~IRS5N2 
(Fig.~\ref{f:map_cd3cn}). Our preliminary LTE analysis of the other two channel
count peak positions, L1551~IRS5N1 and L1551~IRS5S, shows that CD$_3$CN is
not detected toward these positions, which is the reason why we have 
focused the present study on L1551~IRS5N2. Likewise, our preliminary LTE 
models of the channel count peak positions of the other COMPASS sources 
indicates that CD$_3$CN is not detected in the rest of the COMPASS sample.

\subsection{Isotopic ratios and column density of methyl cyanide}
\label{ss:isoratios}

Taken at face value, the column densities reported in Table~\ref{t:coldens}
would imply a $^{12}$C/$^{13}$C isotopic ratio in the range of 12--32, 
depending on the model. This is low compared to the ratio that we expect in 
the local ISM \citep[$\sim$68, with an uncertainty of about 20\%, see the 
discussion in][]{Belloche26}. Likewise, the $^{14}$N/$^{15}$N ratio 
would be in the range of 26--81, much lower than expected for C+N-bearing 
molecules \citep[150--510 in nearby low-mass star forming regions and 250--600 
in high-mass star forming regions at a galactocentric distance of $\sim$8~kpc, 
see][, respectively]{HilyBlant20,Colzi18}. Like for methanol toward IRAS4A2 
\citep[see Sect. 4.1 of][]{Belloche26}, we suspect that our model 
underestimates the actual column density of the main isotopolog of methyl 
cyanide. Therefore,
in the rest of the article, we use $^{13}$CH$_3$CN as a proxy for the column
density of the main isotopolog, multiplying its column density by 68. The last
two columns of Table~\ref{t:coldens} thus list column density ratios that use
$^{13}$CH$_3$CN as a proxy for CH$_3$CN.

\subsection{Uncertainties on column density ratios}
\label{ss:uncertainties}

Table~\ref{t:coldens} shows that, depending on the adopted emission size and
on the molecule, the derived column densities can vary by more than a factor 
of 2. However, in this work we are mostly interested in column density ratios. 
Provided that the emission sizes of two species are the same, their column 
density ratio will be much less uncertain than their individual column 
densities. Following \citet{Belloche26}, we adopt an uncertainty of 15\% on 
the ratios that do not involve a $^{12}$C/$^{13}$C scaling, and 25\% on the 
ratios that do. For ratios that involve CD$_3$CN, we adopt a larger 
uncertainty (25\% on the ratios that do not involve $^{12}$C/$^{13}$C and 30\% 
otherwise), due to the limited number of detected lines and their low 
signal-to-noise ratios.

\input{tab_l1551irs5_p2_cd3cn_deuteration.tex}

\subsection{Deuteration of methyl cyanide}
\label{ss:deuteration}

Table~\ref{t:deuteration} provides column density ratios of various pairs of
the main and deuterated isotopologs of methyl cyanide computed with the
column densities of the best-fit model listed in Table~\ref{t:coldens}.  The
ratios obtained with the alternative models are consistent within the 
uncertainties. We 
accounted for the statistics of the D and H atoms \citep[][]{Faure15} to 
derive D/H ratios from these column density ratios. Table~\ref{t:deuteration}
also lists the D/H ratios.
Following \citet{Faure15}, the relevant equations are: 
$R($CH$_2$DCN$) = 3\alpha$, 
$R($CHD$_2$CN$) = 3\alpha^2$,
and $R($CD$_3$CN$) =\alpha^3$, with $\alpha$ the D/H ratio. The values listed
in the last two rows of Table~\ref{t:deuteration} were obtained by combining
these equations. We note that the values of both $R$ and $\alpha$ used in 
these equations are the decimal numbers, not the percentage values.

\section{Discussion}
\label{s:discussion}

\subsection{Comparison to other sources}
\label{ss:comp_sources}

\begin{figure}[!t]
\sidecaption
\includegraphics[angle=0,width=\hsize]{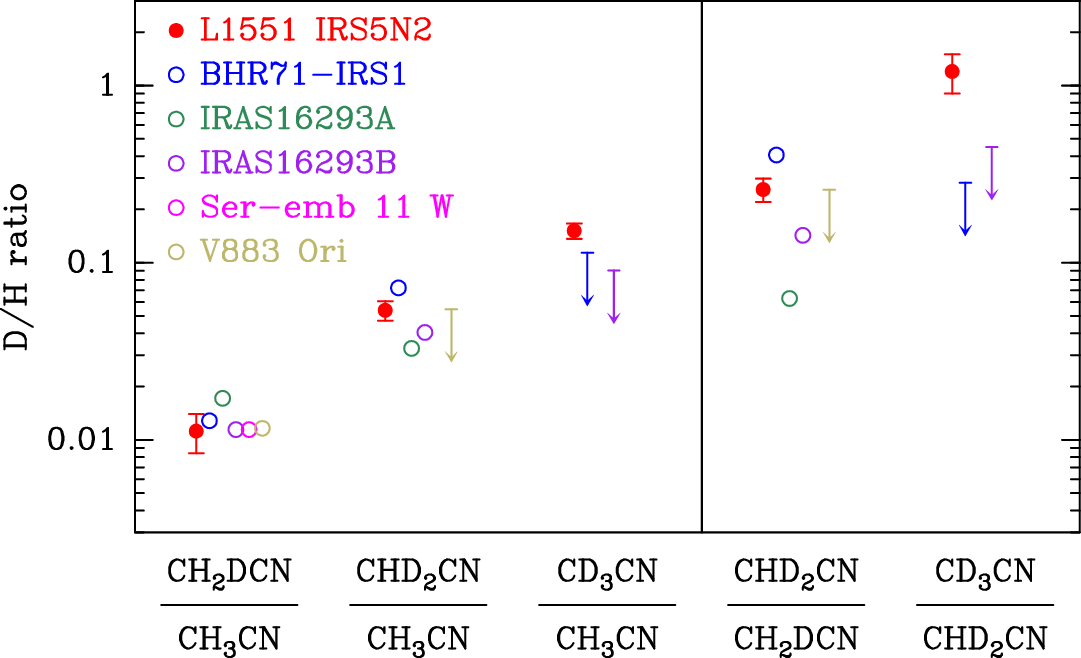}
\caption{Statistically corrected D/H ratios measured toward L1551~IRS5N2 with 
COMPASS and other low-mass protostars from the literature (see 
Sect.~\ref{ss:comp_sources}). The ratios that 
involve CH$_3$CN were computed using the column density of $^{13}$CH$_3$CN 
multiplied by the assumed $^{12}$C/$^{13}$C isotopic ratio.
}
\label{f:deuteration}
\end{figure}

Figure~\ref{f:deuteration} compares the D/H ratios listed in 
Table~\ref{t:deuteration} for L1551~IRS5N2 with results 
reported in the literature for other low-mass protostars. The latter were 
taken from \citet{Nazari26} for BHR71-IRS1, \citet{Calcutt18} for 
IRAS16293--2422A and B (hereafter IRAS16293A and IRAS16293B), 
\citet{MartinDomenech21} for Ser-emb~11~W, and \citet{Jeong25} for 
V883-Ori. In all cases, the ratios involving CH$_3$CN were computed using 
$^{13}$CH$_3$CN as a proxy for CH$_3$CN, with an assumed $^{12}$C/$^{13}$C 
isotopic ratio of 68 (all sources are in the solar neighborhood). We did not 
include the CH$_2$DCN/CH$_3$CN ratio reported by \citet{Bianchi22} toward the 
Class I protostar NGC1333-SVS13A because it relies on single-dish data and an 
uncertain assumption concerning the emission size. In addition to this, these 
authors did not report the column density of the 
$^{13}$C isotopologs, implying a likely overestimated CH$_2$DCN/CH$_3$CN column 
density ratio due to the optical depth of the main isotopolog, as we saw with 
COMPASS for L1551~IRS5N2. 

Figure~\ref{f:deuteration} reveals that the deuteration
pattern of methyl cyanide is similar in all six sources. The D/H ratio of
CH$_2$DCN is almost identical in all sources, with IRAS16293A deviating the
most, although by less than a factor of 1.6. At the next level of deuteration 
(CHD$_2$CN), all sources show a higher D/H ratio than the mono-deuterated
isotopolog, albeit with a larger 
dispersion (by slightly more than a factor of 2). The detection of CD$_3$CN 
toward L1551~IRS5N2 reveals that the trend of increasing D/H ratio with the
multiplicity of deuteration continues, with a D/H increase of about half an 
order of magnitude at each step (left part of Fig.~\ref{f:deuteration}).

On the basis of the COMPASS survey, \citet{Nazari26} reported an upper limit 
for CD$_3$CN toward BHR71-IRS1. We also obtained an upper limit of 
$3\times10^{13}$~cm$^{-2}$ for the column density of CD$_3$CN toward IRAS16293B
from the PILS survey (A. Coutens, priv. comm.). We did not find any CD$_3$CN 
upper limit for any other protostar in the literature. The upper limit to the 
D/H ratio derived by \citet{Nazari26}
from CD$_3$CN/CH$_3$CN is only slightly lower than the ratio we measured toward
L1551~IRS5N2. However, the D/H ratio derived from CHD$_2$CN/CH$_3$CN is slightly
higher for BHR71-IRS1 than for L1551~IRS5N2. This results in a D/H ratio derived
from CD$_3$CN/CHD$_2$CN for L1551~IRS5N2 being much higher (by a factor of 4) 
than the upper limit obtained for BHR71-IRS1. This difference is at the 
3$\sigma$ level (of the uncertainty estimated for L1551~IRS5N2) and may thus be 
significant and indicate an enhanced level of CH$_3$CN deuteration toward 
L1551~IRS5N2 compared to BHR71-IRS1. In the case of IRAS16293B, the D/H ratio
derived from CHD$_2$CN/CH$_3$CN is slightly lower (by $\sim$30\%) than for 
L1551~IRS5N2, and the upper limit to the D/H ratio obtained from 
CD$_3$CN/CH$_3$CN indicates an even slightly more pronounced difference, which 
translates into an upper limit to the D/H ratio derived from 
CD$_3$CN/CHD$_2$CN that is intermediate between that of L1551~IRS5N2 and 
BHR71-IRS1.

\subsection{Deuteration pattern of methyl cyanide and methanol in protostars}
\label{ss:ch3cn_ch3oh}

The deuteration pattern of methyl cyanide displayed in Fig.~\ref{f:deuteration}
is similar to the deuteration pattern of the methyl group of methanol that has
been reported for low-mass protostars, with a D/H ratio increasing from 
CH$_2$DOH to CD$_3$OH (see, e.g., Fig.~5 of 
\citeauthor{Belloche26} \citeyear{Belloche26} and Fig.~4 of 
\citeauthor{Coutens26} \citeyear{Coutens26}). The D/H ratios reported for
methanol in five protostars are on average on the order of 2\% for CH$_2$DOH, 
10\% for CHD$_2$OH, and 25\% for CD$_3$OH. These values are about a factor of 
2 higher than for CH$_3$CN ($\sim$1\% for CH$_2$DCN, $\sim$5\% for CHD$_2$CN, 
and $\sim$15\% for CD$_3$CN), but their progression is similar. This strongly 
suggests that the formation of methyl cyanide has some characteristics in 
common with the formation of methanol. 
Given that it is well established that methanol forms in the solid phase at 
the surface of dust grains during the cold prestellar phase 
\citep[e.g.,][]{Charnley97,Watanabe02,Garrod06,JimenezSerra25,Spezzano26}, it 
is tempting to conclude that the formation of methyl cyanide in low-mass 
protostars is dominated by surface processes during the prestellar phase
\citep[see also][]{Nazari26}. The tight correlation found between the column
densities of methanol and methyl cyanide on the basis of other interferometric 
surveys of protostars also tends to support this conclusion
\citep[][]{Belloche20,Yang21}.

\subsection{Comparison to astrochemical models}

Two astrochemical models dealing with the deuteration of COMs have 
investigated the chemical evolution of collapsing low-mass dense cores from the 
prestellar phase to the end of the Class 0 protostellar phase 
\citep[][]{Taquet12,Taquet14, Aikawa12}. These models couple the chemistry in 
the gas phase and on the surface and in the ice mantles of dust grains. The
model of \citet[][, hereafter T14]{Taquet14} accounts for the spin state of 
H$_2$ (ortho and para) and for the layering of the grain mantles, which is not 
the case for the model of \citet[][, hereafter A12]{Aikawa12}. Model A12 
consists of two stages:
a cold, hydrostatic stage lasting for $10^6$~yr followed by a phase of 
gravitational collapse described by a one-dimensional, radiation-hydrodynamic,
numerical model, while model T14 uses a parametric description of the 
quasistatic contraction of a dense core, followed by a free-fall 
collapse phase that quickly leads to the formation of a protostar. The dust 
and gas temperatures are kept equal in both models, but their profiles
differ: the temperature drops from 17~K to 7~K from the outer to the inner
parts of the dense core at the end of the prestellar phase in model T14 while
the external temperature is 10~K and the inner temperature reaches 5~K at the
end of the prestellar phase in model A12. The temperature increases in both 
models once the protostellar embryo has formed.

The chemical networks of both models include the formation of CH$_2$DCN but, 
unfortunately, none of them deals with CHD$_2$CN and CD$_3$CN. Their 
predictions of the gas-phase CH$_2$DCN/CH$_3$CN abundance ratio differ 
significantly, which may be due to the different treatments of the mantles
(multilayer in T14 versus homogeneous in A12): the ratio drops from 3.6\% to 
0.15\% during the Class 0 phase in model T14 while it increases from 2.3\% to
6.3\% in model A12. The column density ratio measured toward L1551~IRS5N2 
(3.4\%) is consistent with the ranges predicted by both models. The inclusion 
of CHD$_2$CN and CD$_3$CN in the networks of both models might help us in the 
future to discriminate them. \citet{Taquet14} plotted the gas-phase abundance 
profiles of COMs thought to be mainly formed on the grains in their Fig. 11. 
The figure includes methyl cyanide, suggesting that its formation is dominated 
by solid-phase processes in model T14. These authors mention two formation 
pathways on the grains: the hydrogenation of C$_2$N during the cold phase and 
the radical recombination of CH$_3$ and CN during the warm-up phase. 
\citet{Aikawa12} also mention the cold formation of CH$_3$CN on the grains 
(addition of C and H to CN by association reactions), and their Fig. 2 
indicates that the abundance of CH$_3$CN in the ice continues to increase (by 
a factor of a few) during the warm-up phase.

Another view on the formation pathways of methyl cyanide can be found in 
\citet{Garrod22}. These authors investigated the chemical evolution of 
star forming regions with a three-phase (gas, surface and bulk ice mantles 
of dust grains) astrochemical code. One of the strengths of this work is the 
inclusion of nondiffusive processes in the chemical network. The simulations 
involve a two-stage physical treatment: a cold, collapse phase from a density
of $3 \times 10^3$~cm$^{-3}$ to $2 \times 10^8$~cm$^{-3}$ followed by a warm-up
phase from 8 to 400~K at fixed density\footnote{The density at which 
the temperature reaches 100~K in L1551~IRS5 happens to be on the order of 
$2 \times 10^8$~cm$^{-3}$ as well \citep[see Table 2 of][]{Jorgensen26}. The 
conditions considered in the models of \citet{Garrod22} are thus applicable to 
this source.}. Three warm-up timescales, called fast, 
medium, and slow, were investigated: $5 \times 10^4$~yr, $2 \times 10^5$~yr, 
and $1 \times 10^6$~yr, respectively. These timescales correspond to the time 
when the temperature reaches 200~K. In the final model with the medium 
timescale, about 90\% of the formation of methyl cyanide occurs in the gas 
phase during the warm-up phase at temperatures above 100~K (reaction of 
CH$_3^+$ with HCN, followed by electronic recombination, see their Fig.~14), 
while the solid-phase chemistry during the cold stage only accounts for about 
10\%. In the fast timescale model, the proportions are about 50:50 (see their 
Fig.~19). 

The lifetimes of Class~0 and I protostars is on the order of 
$1 \times 10^5$~yr and $5 \times 10^5$~yr, respectively 
\citep[][]{Evans09,Enoch09}. Therefore, the Class I protostar L1551~IRS5 
is at least $1 \times 10^5$~yr old. While this might at first sight suggest 
that the model with the fast timescale is not appropriate for 
this source, what actually matters is the time it took for the material 
currently observed in the hot corino region to reach this region from the
place it originally occupied when it started to feel the heating from the
protostar. This is not straightforward to estimate but
3D magnetohydrodynamical (MHD) simulations of the formation of protostars in a 
dynamic molecular cloud environment have for instance predicted
timescales shorter than $10^4$~yr for a test particle to feel a 
sharp temperature increase exceeding 100~K during the collapse
\citep[see, e.g., Fig.~9 of][]{Jensen21}. Therefore, we do not exclude any of 
the two models (fast or medium timescale) of \citet{Garrod22} for our 
comparison to L1551~IRS5.

In Sect.~\ref{ss:ch3cn_ch3oh}, we emphasized that the progression of the 
deuteration levels of methyl cyanide and methanol is similar from one 
deuterated isotopolog to the next. However, a key point is that their D/H 
ratios overall differ by a factor of $\sim$2. This could imply that methyl 
cyanide and its deuterated isotopologs were formed during the prestellar phase 
at a slightly warmer, hence earlier, stage than methanol. Alternatively, the 
lower D/H ratios of methyl cyanide could result from a gas-phase contribution 
to the production of methyl cyanide during the (deuteration-inefficient) 
warm-up stage, in addition to the initial solid-state formation of methyl 
cyanide and its deuterated isotopologs during the cold prestellar phase. In 
the latter case, the factor of $\sim$2 mentioned above would then point to 
similar contributions from a (cold and deuteration-efficient) solid-phase 
pathway and a (warm and deuteration-inefficient) gas-phase pathway to the 
production of methyl cyanide. The latter scenario would be in line with the 
predictions of the fast-timescale model of \citet{Garrod22}. However, the 
efficient gas-phase production of methyl cyanide in this model only occurs 
above 100~K and requires about $5 \times 10^4$~yr to happen, which is much 
longer than the aforementioned timescale of the MHD simulations discussed in 
\citet{Jensen21}. Therefore, whether gas-phase chemistry can significantly 
contribute to the production of methyl cyanide in L1551~IRS5N in the context 
of the model of \citet{Garrod22} is not clear. Unfortunately, the chemical 
network of this model does not include deuteration and we cannot compare the 
deuteration pattern of methyl cyanide measured toward L1551~IRS5N2 to 
predictions of this model. 

While the MHD simulations discussed in \citet{Jensen21} provide valuable 
insight into the large-scale dynamics of the system, they do not resolve the 
formation of an accretion disk, nor were they intended to reproduce any 
specific system in detail. L1551~IRS5 is known to exhibit a distinctive 
small-scale structure comprising two circumstellar disks (CSD) around the 
binary components N and S and a circumbinary disk 
\citep[CBD,][]{CruzSaenzDeMiera19}. This CBD consists of two spiral arms that 
connect the circumbinary structure to each CSD \citep[][]{Takakuwa20}. In the 
latter study, both the structure and kinematics of these components were well 
fitted with the 3D HD adaptive-mesh-refinement numerical simulations of 
\citet{Matsumoto19} with a binary mass ratio of unity. In the framework of
this model, each CSD is confined within the Roche lobe of its protostar, and
the first Lagrangian point, L1, is located half-way between N and S, at 
0.2$\arcsec$ from each protostar. This means that the position where CD$_3$CN
is detected, N2, is located approximately half-way between L1 and N, where the
dynamics seems to be dominated by rotation in the simulations of 
\citet{Matsumoto19} (see their Fig.~2). The dynamical timescale associated 
with the parcel of gas where CD$_3$CN is detected may thus be longer than 
the infall timescale inferred from the MHD simulations discussed in 
\citet{Jensen21}, giving gas-phase chemistry more time to contribute to the
production of CH$_3$CN after sublimation from the grain mantles.
However, a recent study of L1551-IRS5 at high angular resolution 
suggests that COM abundances in the region near N2 may be enhanced by shocks 
produced by an infalling envelope streamer 
traced by SO \citep[][]{Kang26}. The timescale relevant to the chemistry might
then be related to the shocks rather than to the CSD orbital time.
Coupling astrochemical simulations to 3D MHD simulations that account for the 
small-scale architecture of L1551~IRS5 would allow us to understand further 
how the dynamical evolution influences the chemical composition of star 
forming regions.

\section{Conclusions}
\label{s:conclusions}

Thanks to the high sensitivity of the ALMA data set obtained toward L1551~IRS5
in the framework of \hbox{COMPASS}, we have detected all 
deuterated isotopologs of methyl cyanide toward the northern component of this
binary system. In particular, we report the first interstellar detection of
fully deuterated methyl cyanide, CD$_3$CN. The deuteration pattern measured for
methyl cyanide, with an increase in the D/H ratio by about half an order of 
magnitude at each step of deuteration, is similar to the deuteration pattern
measured for methanol in several other low-mass protostars. This
suggests that methyl cyanide predominantly forms like methanol on the surface 
of dust grains at low temperature. However, the overall factor of $\sim$2
difference between the D/H ratios of methyl cyanide and methanol may either 
indicate that gas-phase formation of methyl cyanide during the
(deuteration-inefficient) warm-up
stage is approximately as important as its cold, solid-state formation, or 
that methyl cyanide and its deuterated isotopologs are formed somewhat earlier 
than methanol during the prestellar phase.

Unfortunately, no astrochemical modeling studies have investigated the 
deuteration of methyl cyanide up to its fully deuterated form so far. The 
models of \citet{Taquet14} and \citet{Aikawa12} predict a CH$_2$DCN/CH$_3$CN
abundance ratio consistent with the column density ratio measured toward 
L1551~IRS5, but no predictions are available for CHD$_2$CN and CD$_3$CN. 
In the astrochemical model of \citet{Garrod22}, the formation of methyl cyanide 
occurs both during the cold prestellar phase in the solid state and during
the warm-up phase at temperatures above 100~K in the gas. The balance between 
these two pathways depends on the warm-up timescale, with a predominance of the 
gas-phase pathway in the medium-timescale model and similar contributions of 
both pathways in the fast-timescale model. The latter scenario appears to be in 
line with the observational results mentioned above, which could suggest that 
the most appropriate evolution of physical conditions in L1551~IRS5 is that 
described by the fast-timescale model. Further progress in our understanding 
of the formation of methyl cyanide in the ISM could be made with dedicated 
astrochemical simulations that are tailored for low-mass protostars and 
include all three deuterated isotopologs of this molecule in their networks.


\begin{acknowledgements}
This paper makes use of the following ALMA data: ADS/JAO.ALMA\#2022.1.00316.L. 
ALMA is a partnership of ESO (representing its member states), NSF (USA) and 
NINS (Japan), together with NRC (Canada), NSTC and ASIAA (Taiwan), and KASI 
(Republic of Korea), in cooperation with the Republic of Chile. The Joint ALMA 
Observatory is operated by ESO, AUI/NRAO, and NAOJ.
The National Radio Astronomy Observatory is a facility of the National Science
Foundation operated under cooperative agreement by Associated Universities,
Inc.
The project that led to this publication has received support from ORP that is 
funded by the European Union's Horizon 2020 research and innovation programme 
under grant agreement No 101004719 [ORP]. 
We acknowledge financial support from the following: 
the Danish National Research Foundation through the Center of Excellence 
``InterCat'' (Grant agreement no.: DNRF150; J.K.J.);
the European Research Council (ERC) under the European Union's Horizon 2020  
research and innovation programme (ERC Starting Grant 
``Chemtrip'', grant agreement No 949278; A.C., F.C.-S.M.);
the Max Planck Society (S.S.); 
the National Research Foundation of Korea (NRF), funded by the Ministry of 
Science and ICT (MSIT) under grant numbers RS-2024-00416859 and 
RS-2026-25490557 (J.-E.L., J.-H.J., H.-S.Y.), and by the Ministry of Education 
through the Basic Science Research Program under grant number RS-2025-25402622 
(J.-H.J.); 
the Ministry of Education of Taiwan (a grant from Center for Informatics and 
Computation in Astronomy, CICA, and grant number 110J0353I9) and the National 
Science and Technology Council, Taiwan (Grant NSTC111-2112-M-007-014-MY3, 
NSTC113-2639-M-A49-002-ASP, and NSTC113-2112-M-007-027; D.H.); 
the Thematic Action ``Physique et Chimie du Milieu Interstellaire'' (PCMI) of 
INSU Programme National ``Astro'', with contributions from CNRS Physique \& 
CNRS Chimie, CEA, and CNES (S.M.); 
the Ministry of Education, Culture, Sports, Science, and Technology of Japan
(Grant-in-Aid: 25H00676) and RIKEN (Incentive Research Project "Emergence of 
pre-biotic extraterrestrial world: Using machine-learning techniques to 
uncover the link of organic matter at the dawn of planetary systems") 
(Y.-L.Y.). P.N. is supported by a Warwick Prize Fellowship.
MND is funded by the European Union. Views and opinions expressed are however 
those of the author(s) only and do not necessarily reflect those of the 
European Union or the European Research Council Executive Agency. Neither the 
European Union nor the granting authority can be held responsible for them. 
MND's work is supported by ERC grant PSII (DOI: 10.3030/101230593).
\end{acknowledgements}

\bibliographystyle{aa}
\bibliography{compass_l1551irs5_cd3cn}

\begin{appendix}
\label{appendix}

\onecolumn
\section{Spectra}
\label{a:spectra}

\begin{figure*}[!h]
\centerline{\resizebox{0.95\hsize}{!}{\includegraphics[angle=0]{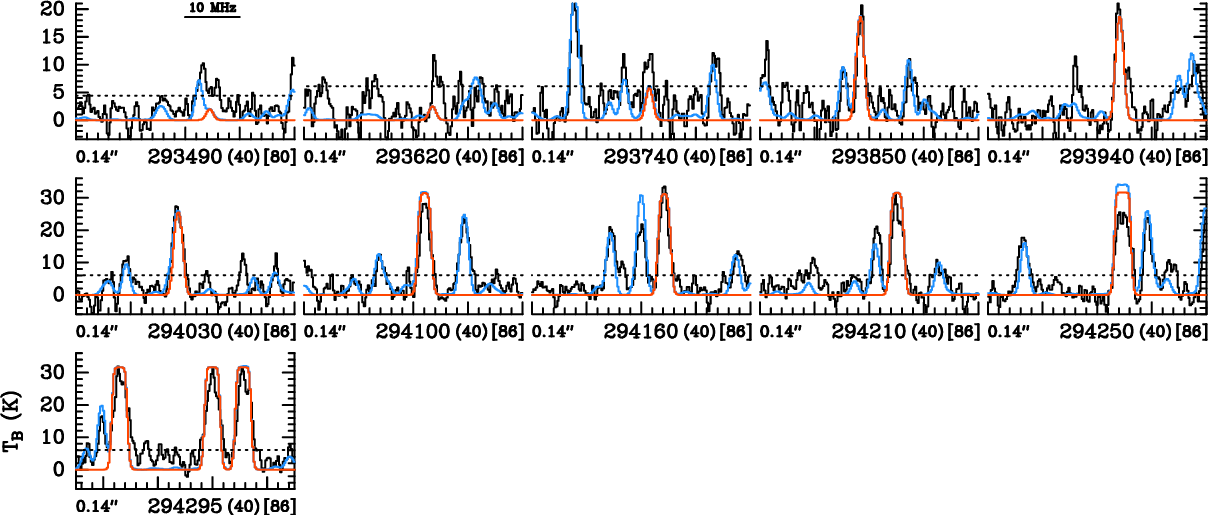}}}
\caption{Transitions of CH$_3$CN $\varv=0$ covered by the COMPASS survey
toward L1551~IRS5N2 with ALMA. The best-fit LTE synthetic spectrum of 
CH$_3$CN $\varv=0$ is shown in red and overlaid on the observed 
spectrum displayed in black. The blue synthetic spectrum contains the 
contribution of all molecules included in our model of L1551~IRS5N2 so far, 
including the molecule shown in red. The 
central frequency is indicated in MHz below each panel as well as the 
HPBW (geometrical mean) on the left, the width of each panel in MHz in 
parentheses, and the continuum level in K of the baseline-subtracted spectra 
in brackets. For guidance, the frequency interval between two big
tickmarks is indicated in the top left panel.
The y-axis is labeled in brightness temperature units (K). The 
dotted line indicates the 3$\sigma$ noise level.}
\label{f:spec_ch3cn_ve0}
\end{figure*}

\begin{figure*}
\centerline{\resizebox{0.95\hsize}{!}{\includegraphics[angle=0]{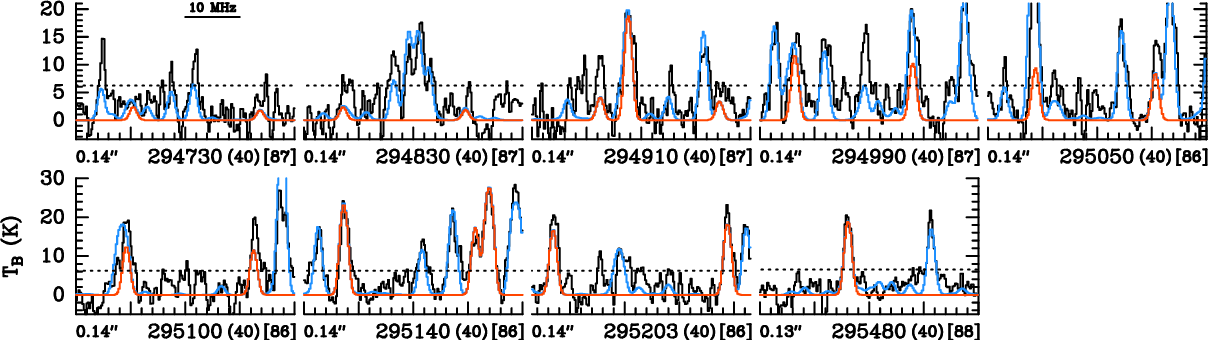}}}
\caption{Same as Fig.~\ref{f:spec_ch3cn_ve0}, but for CH$_3$CN, $\varv_8=1$.}
\label{f:spec_ch3cn_v8e1}
\end{figure*}

\begin{figure*}
\centerline{\resizebox{0.95\hsize}{!}{\includegraphics[angle=0]{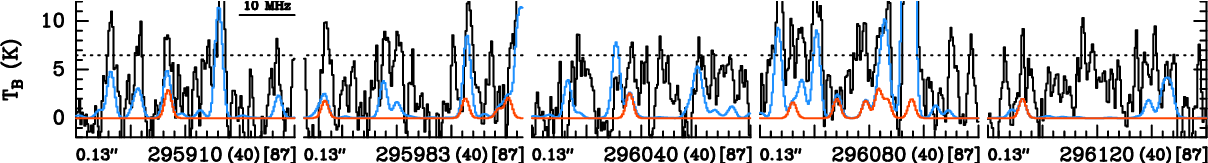}}}
\caption{Same as Fig.~\ref{f:spec_ch3cn_ve0}, but for CH$_3$CN, $\varv_8=2$.}
\label{f:spec_ch3cn_v8e2}
\end{figure*}

\begin{figure*}
\centerline{\resizebox{0.95\hsize}{!}{\includegraphics[angle=0]{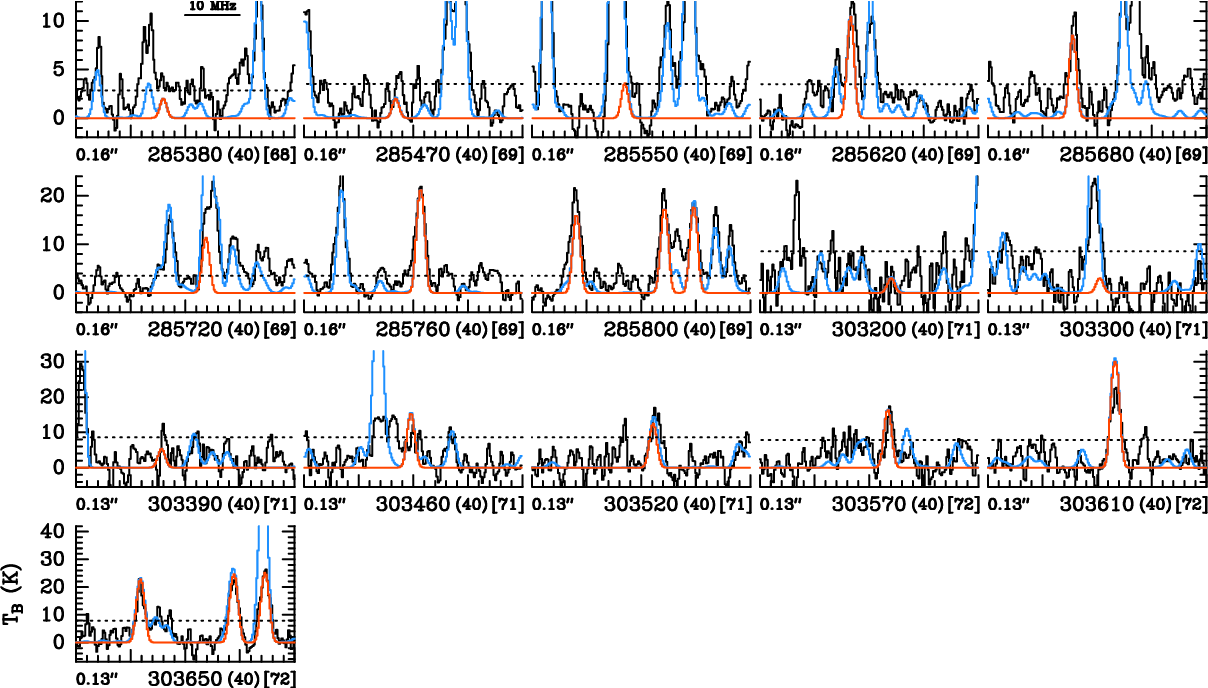}}}
\caption{Same as Fig.~\ref{f:spec_ch3cn_ve0}, but for $^{13}$CH$_3$CN, 
$\varv=0$.}
\label{f:spec_ch3cn_13c1_ve0}
\end{figure*}

\begin{figure*}
\centerline{\resizebox{0.95\hsize}{!}{\includegraphics[angle=0]{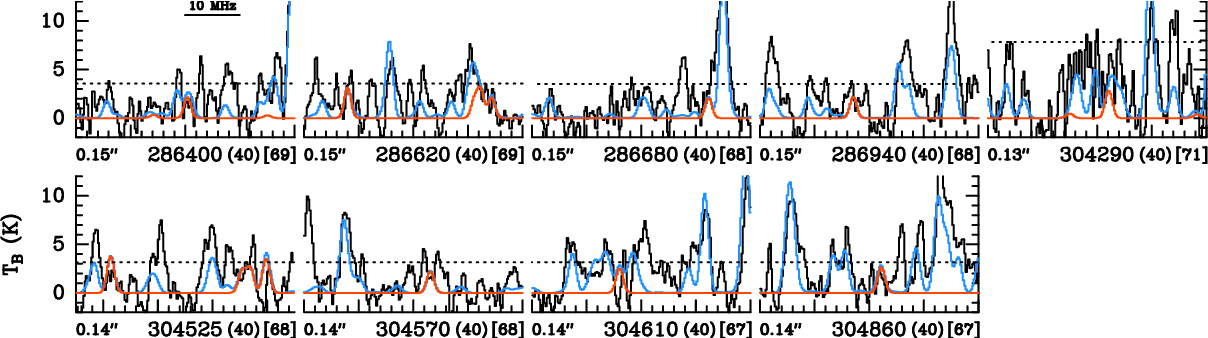}}}
\caption{Same as Fig.~\ref{f:spec_ch3cn_ve0}, but for $^{13}$CH$_3$CN, 
$\varv_8=1$.}
\label{f:spec_ch3cn_13c1_v8e1}
\end{figure*}

\begin{figure*}
\centerline{\resizebox{0.95\hsize}{!}{\includegraphics[angle=0]{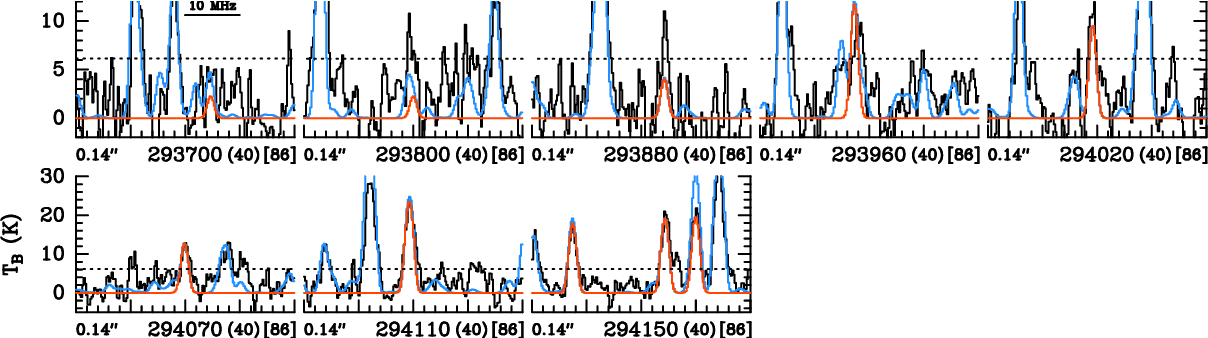}}}
\caption{Same as Fig.~\ref{f:spec_ch3cn_ve0}, but for CH$_3$$^{13}$CN, 
$\varv=0$.}
\label{f:spec_ch3cn_13c2_ve0}
\end{figure*}

\begin{figure*}
\centerline{\resizebox{0.95\hsize}{!}{\includegraphics[angle=0]{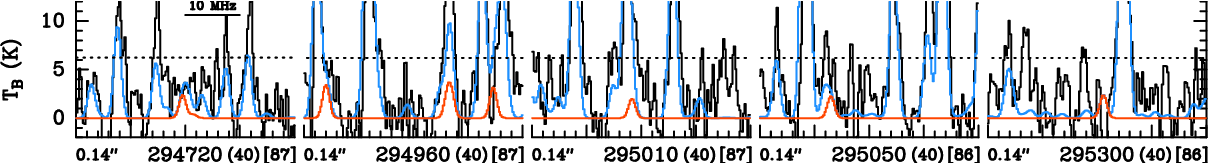}}}
\caption{Same as Fig.~\ref{f:spec_ch3cn_ve0}, but for CH$_3$$^{13}$CN, 
$\varv_8=1$.}
\label{f:spec_ch3cn_13c2_v8e1}
\end{figure*}

\begin{figure*}
\centerline{\resizebox{0.95\hsize}{!}{\includegraphics[angle=0]{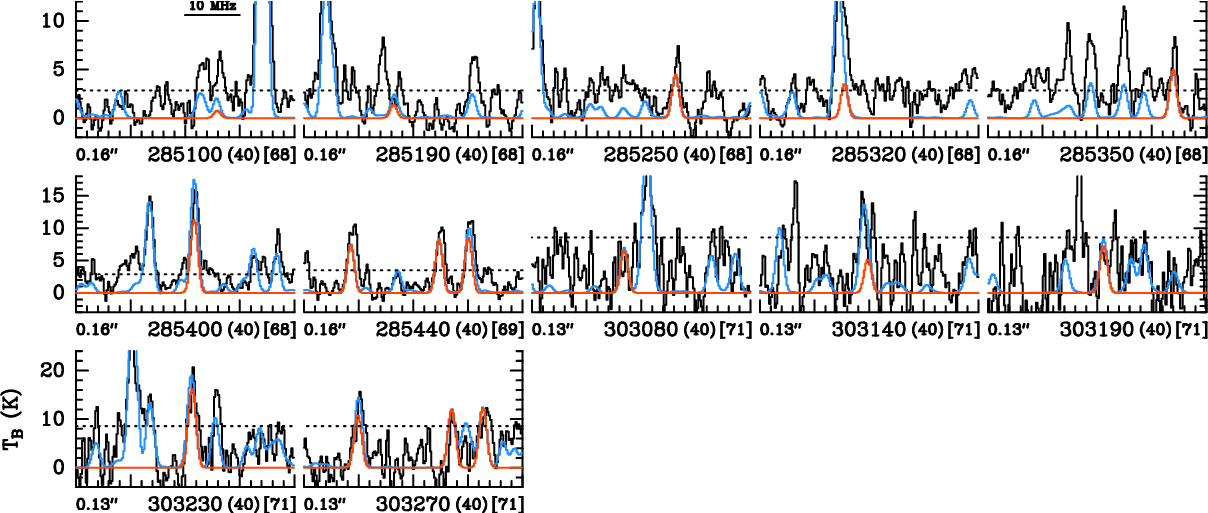}}}
\caption{Same as Fig.~\ref{f:spec_ch3cn_ve0}, but for CH$_3$C$^{15}$N, 
$\varv=0$.}
\label{f:spec_ch3cn_15n_ve0}
\end{figure*}

\begin{figure*}
\centerline{\resizebox{0.41\hsize}{!}{\includegraphics[angle=0]{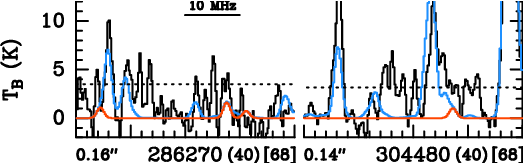}}}
\caption{Same as Fig.~\ref{f:spec_ch3cn_ve0}, but for CH$_3$C$^{15}$N, 
$\varv_8=1$.}
\label{f:spec_ch3cn_15n_v8e1}
\end{figure*}

\begin{figure*}
\centerline{\resizebox{0.95\hsize}{!}{\includegraphics[angle=0]{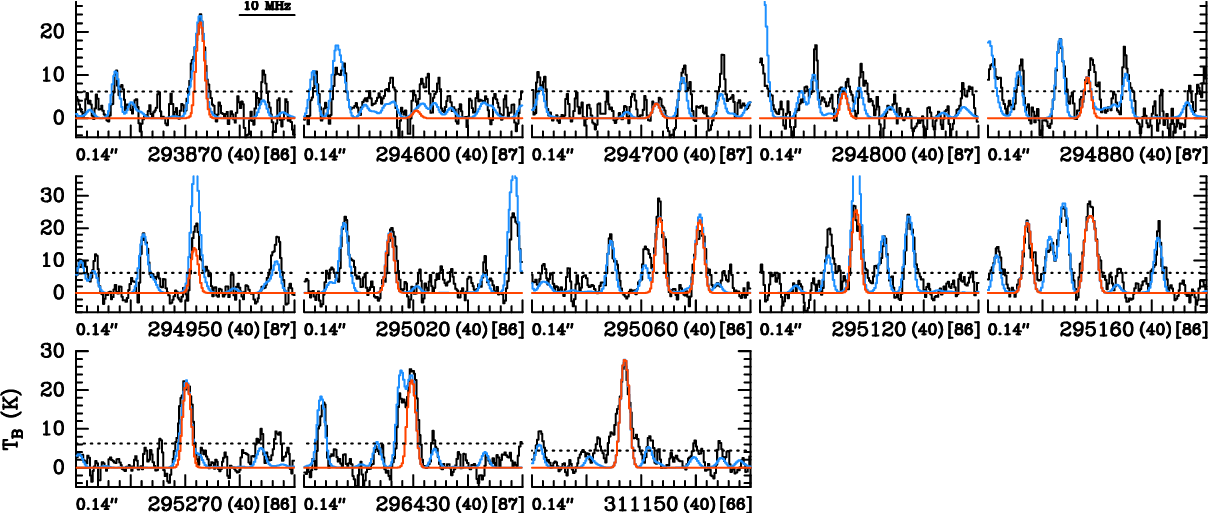}}}
\caption{Same as Fig.~\ref{f:spec_ch3cn_ve0}, but for CH$_2$DCN, 
$\varv=0$.}
\label{f:spec_ch2dcn_ve0}
\end{figure*}

\begin{figure*}
\centerline{\resizebox{0.95\hsize}{!}{\includegraphics[angle=0]{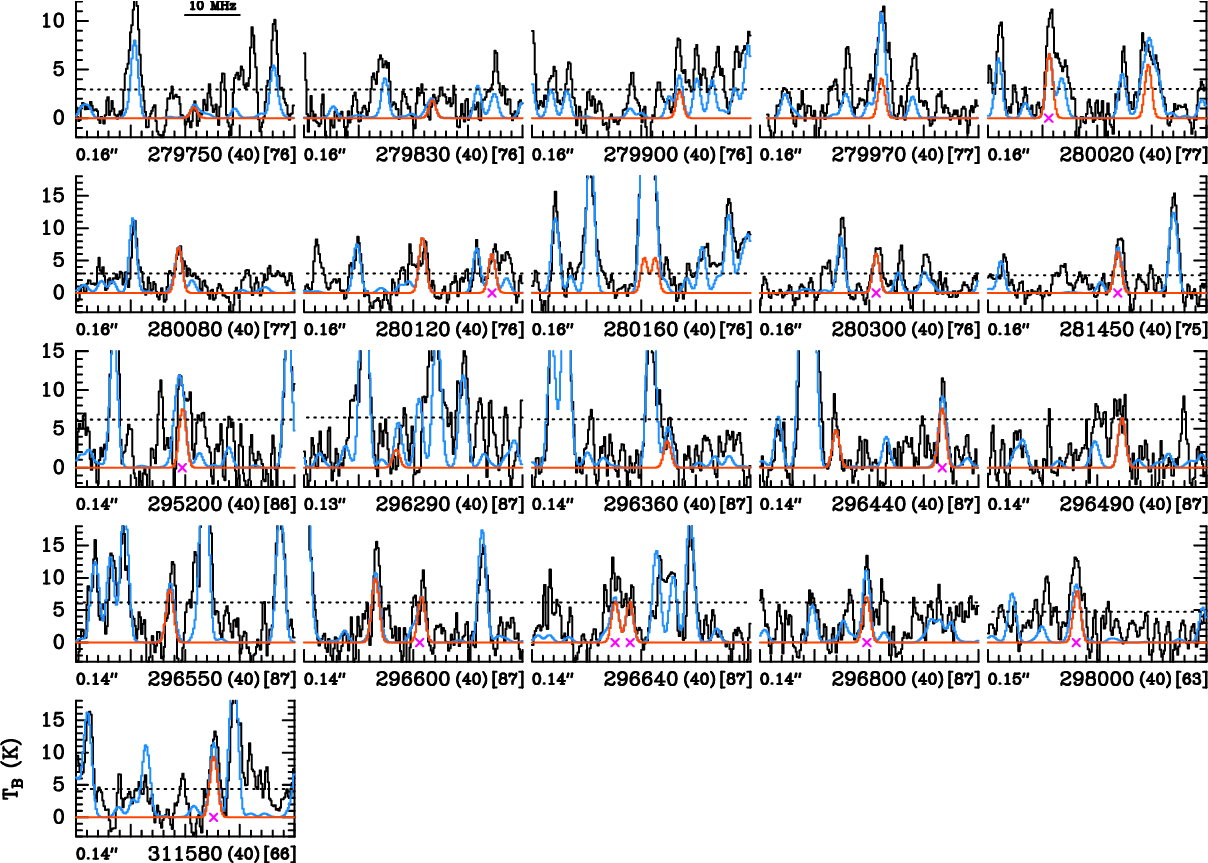}}}
\caption{Same as Fig.~\ref{f:spec_ch3cn_ve0}, but for CHD$_2$CN, 
$\varv=0$. The transitions with an upper-level energy between 120~K and 180~K
in temperature unit are marked with a magenta cross (see Sect.~\ref{ss:param}).}
\label{f:spec_chd2cn_ve0}
\end{figure*}

\begin{figure*}
\centerline{\resizebox{0.95\hsize}{!}{\includegraphics[angle=0]{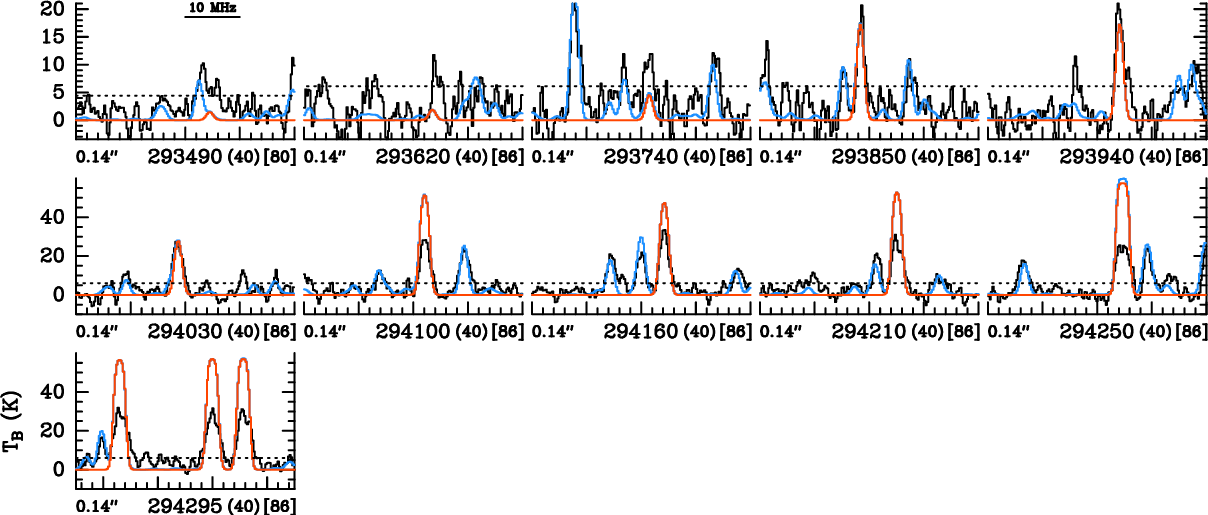}}}
\caption{Same as Fig.~\ref{f:spec_ch3cn_ve0}, but for the alternative model 
with a size of 0.15$\arcsec$ and a temperature of 200~K.}
\label{f:spec_ch3cn_ve0_alt1}
\end{figure*}

\begin{figure*}
\centerline{\resizebox{0.92\hsize}{!}{\includegraphics[angle=0]{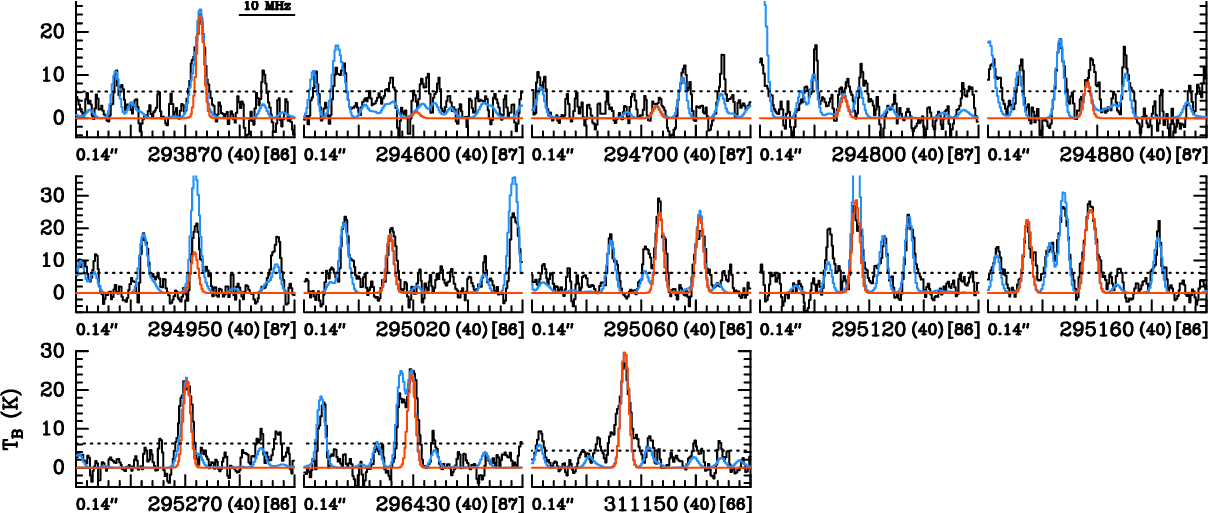}}}
\caption{Same as Fig.~\ref{f:spec_ch3cn_ve0_alt1}, but for CH$_2$DCN, 
$\varv = 0$.}
\label{f:spec_ch2dcn_ve0_alt1}
\end{figure*}

\begin{figure*}
\centerline{\resizebox{0.92\hsize}{!}{\includegraphics[angle=0]{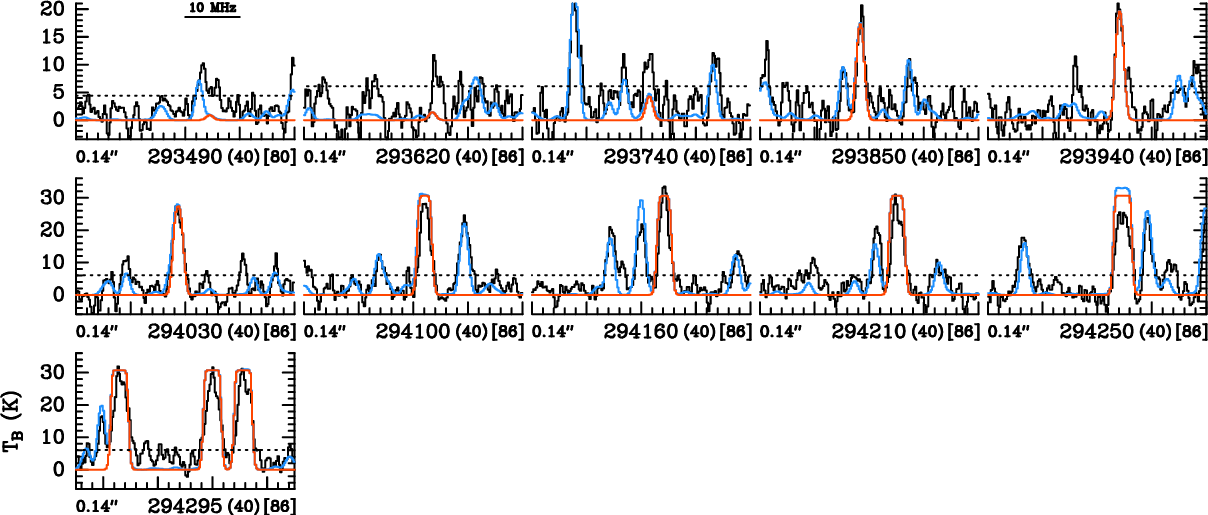}}}
\caption{Same as Fig.~\ref{f:spec_ch3cn_ve0}, but for the alternative model 
with a size of 0.15$\arcsec$ and a temperature of 150~K.}
\label{f:spec_ch3cn_ve0_alt2}
\end{figure*}

\begin{figure*}
\centerline{\resizebox{0.92\hsize}{!}{\includegraphics[angle=0]{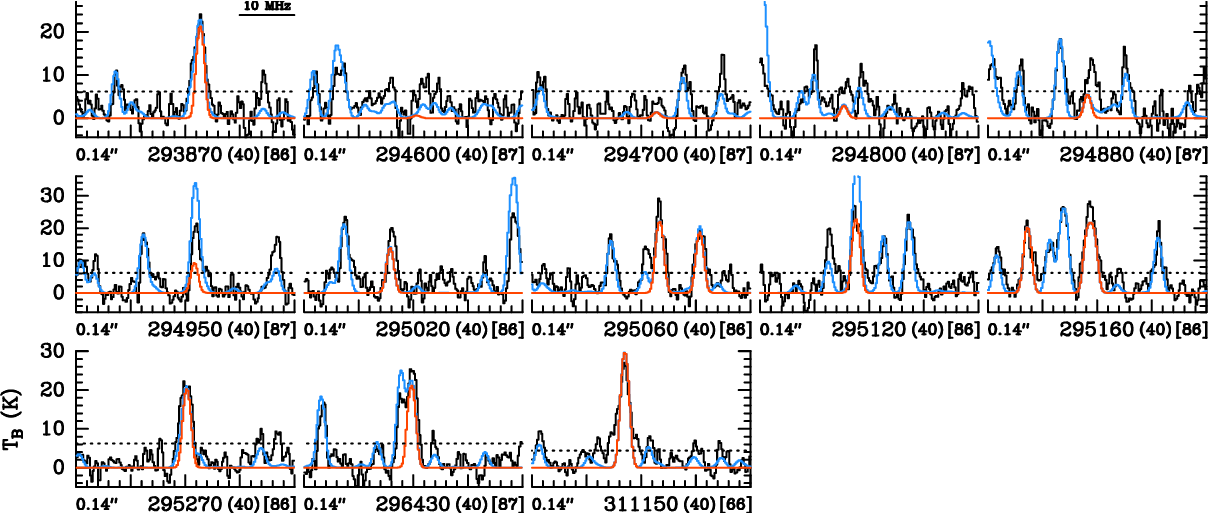}}}
\caption{Same as Fig.~\ref{f:spec_ch3cn_ve0_alt2}, but for CH$_2$DCN, 
$\varv = 0$.}
\label{f:spec_ch2dcn_ve0_alt2}
\end{figure*}

Figures~\ref{f:spec_ch3cn_ve0}--\ref{f:spec_chd2cn_ve0} show the transitions of
CH$_3$CN $\varv=0$, CH$_3$CN $\varv_8=1$, CH$_3$CN $\varv_8=2$, 
$^{13}$CH$_3$CN $\varv=0$, $^{13}$CH$_3$CN $\varv_8=1$,
CH$_3$$^{13}$CN $\varv=0$, CH$_3$$^{13}$CN $\varv_8=1$,
CH$_3$C$^{15}$N $\varv=0$, CH$_3$C$^{15}$N $\varv_8=1$, CH$_2$DCN $\varv=0$,
and CHD$_2$CN $\varv=0$, respectively, that are covered by the COMPASS survey 
and significantly contribute to the signal detected toward L1551~IRS5N2. For 
each isotopolog the transitions that are heavily blended with much stronger 
emission from other species and therefore cannot contribute to the 
identification of that isotopolog are not displayed in these figures.
Figures~\ref{f:spec_ch3cn_ve0_alt1} and \ref{f:spec_ch2dcn_ve0_alt1} show the
spectra of CH$_3$CN $\varv=0$ and CH$_2$DCN $\varv=0$ for the first alternative
model and Figs.~\ref{f:spec_ch3cn_ve0_alt2} and 
\ref{f:spec_ch2dcn_ve0_alt2} for the second alternative model.

\clearpage
\twocolumn
\section{Population diagrams}
\label{a:popdiag}

\begin{figure}[!h]
\centerline{\resizebox{0.80\hsize}{!}{\includegraphics[angle=0]{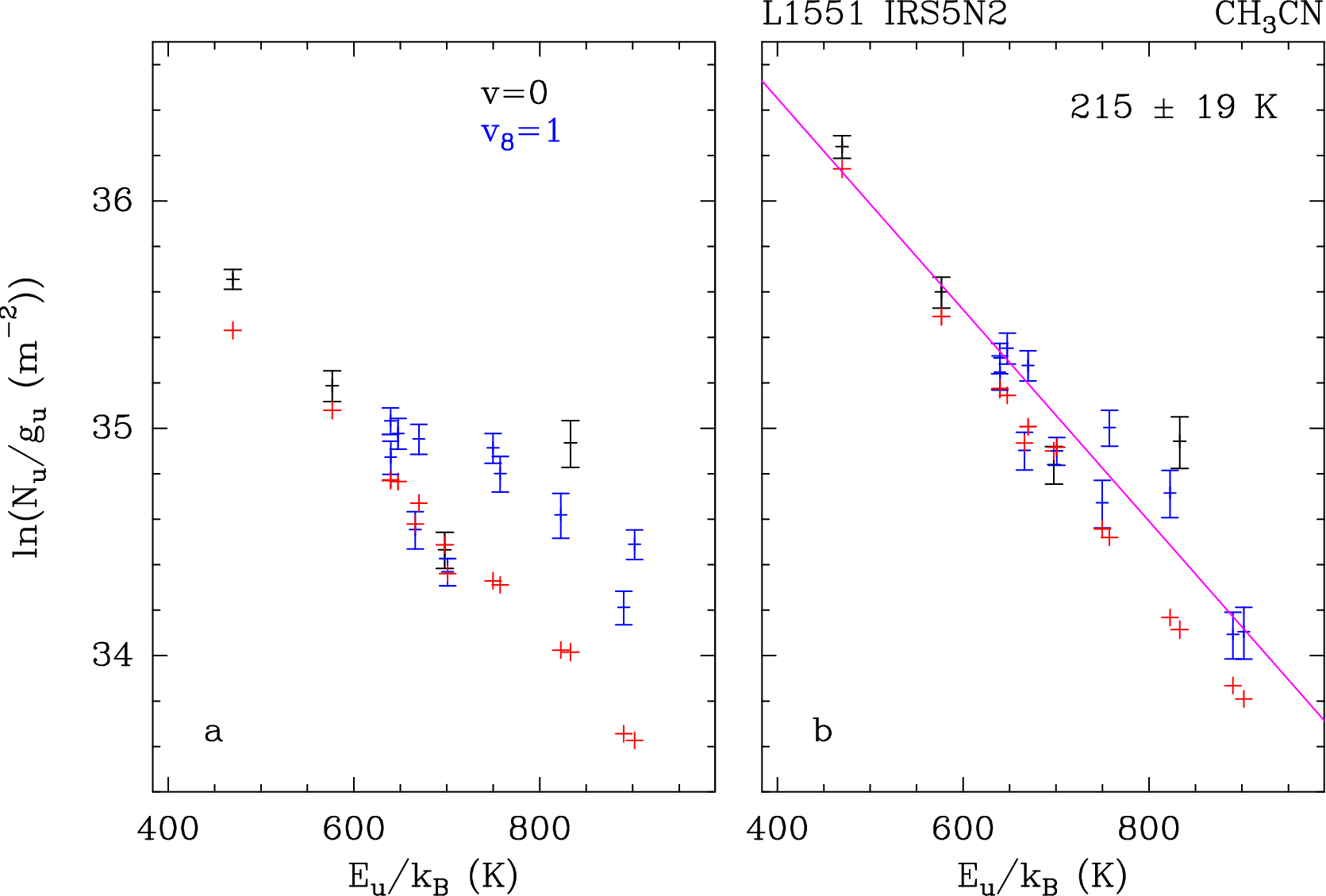}}}
\vspace*{-1ex}
\caption{Population diagrams of CH$_3$CN toward L1551~IRS5N2. The observed 
data points are shown in various colors (corresponding to the vibrational 
states), except for red, as indicated in the upper right corner of panel a 
while the synthetic populations of the best-fit Weeds model are shown in red. 
No correction is applied in panel a. In panel b, the optical depth 
correction has been applied to both the observed and synthetic populations and 
the contamination by all other species included in the full model has been 
removed from the observed data points. The purple line is a linear fit to the 
observed populations and the rotational 
temperature derived from this fit is indicated in the top right corner.}
\label{f:popdiag_ch3cn}
\end{figure}

\begin{figure}[!h]
\centerline{\resizebox{0.80\hsize}{!}{\includegraphics[angle=0]{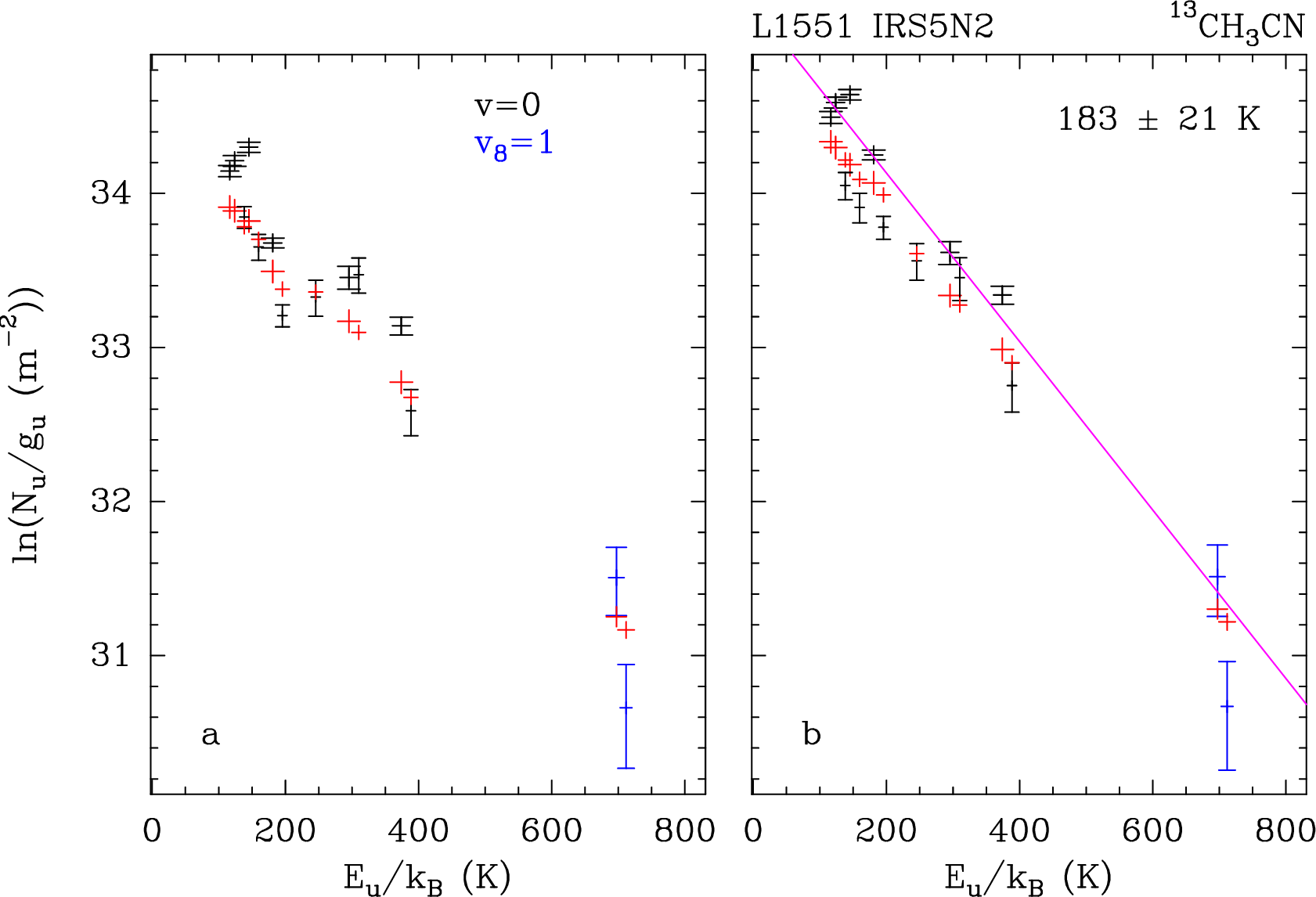}}}
\vspace*{-1ex}
\caption{Same as Fig.~\ref{f:popdiag_ch3cn}, but for $^{13}$CH$_3$CN.}
\label{f:popdiag_ch3cn_13c1}
\end{figure}

\begin{figure}[!h]
\centerline{\resizebox{0.80\hsize}{!}{\includegraphics[angle=0]{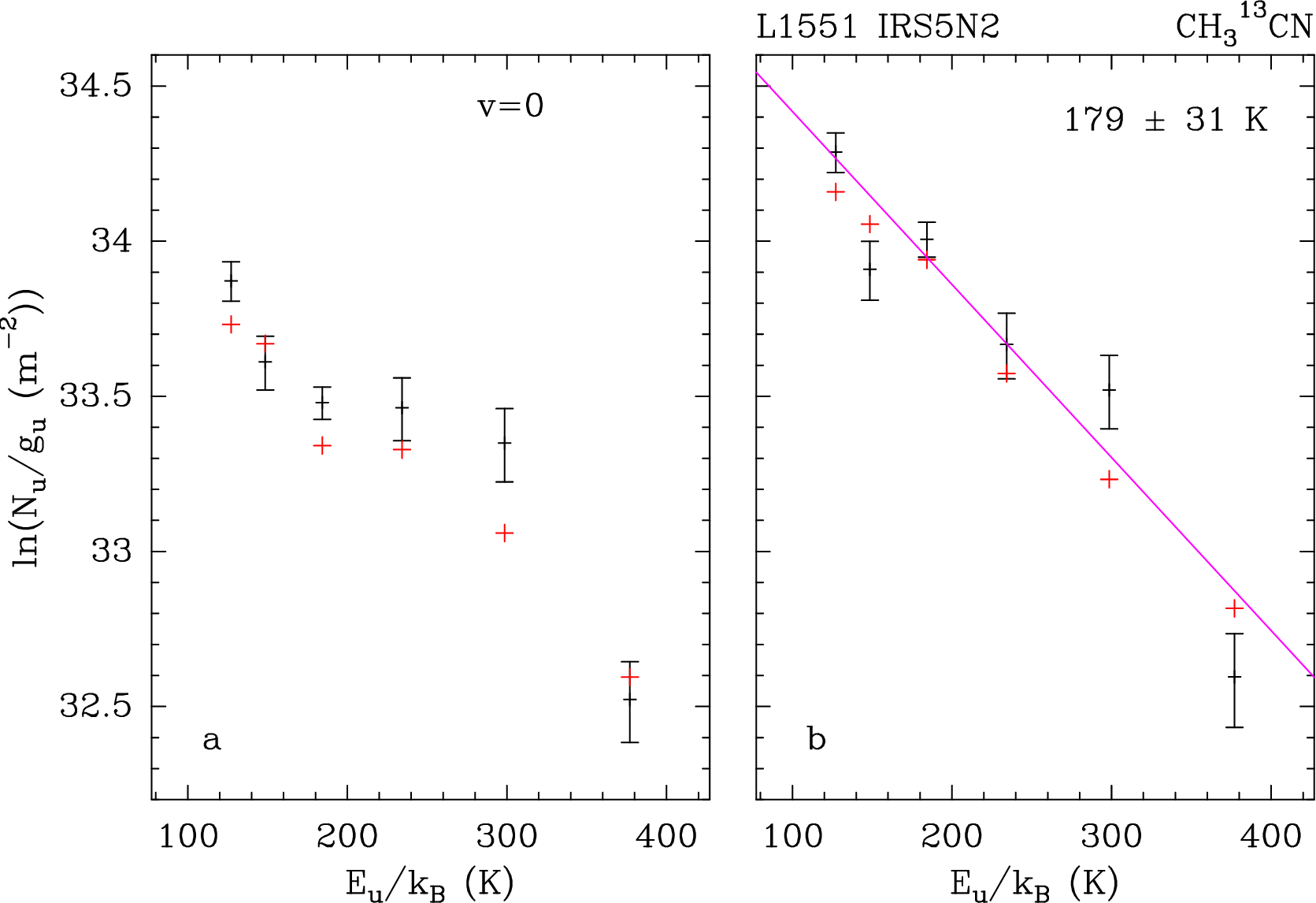}}}
\vspace*{-1ex}
\caption{Same as Fig.~\ref{f:popdiag_ch3cn}, but for CH$_3$$^{13}$CN.}
\label{f:popdiag_ch3cn_13c2}
\end{figure}

\begin{figure}[!h]
\centerline{\resizebox{0.80\hsize}{!}{\includegraphics[angle=0]{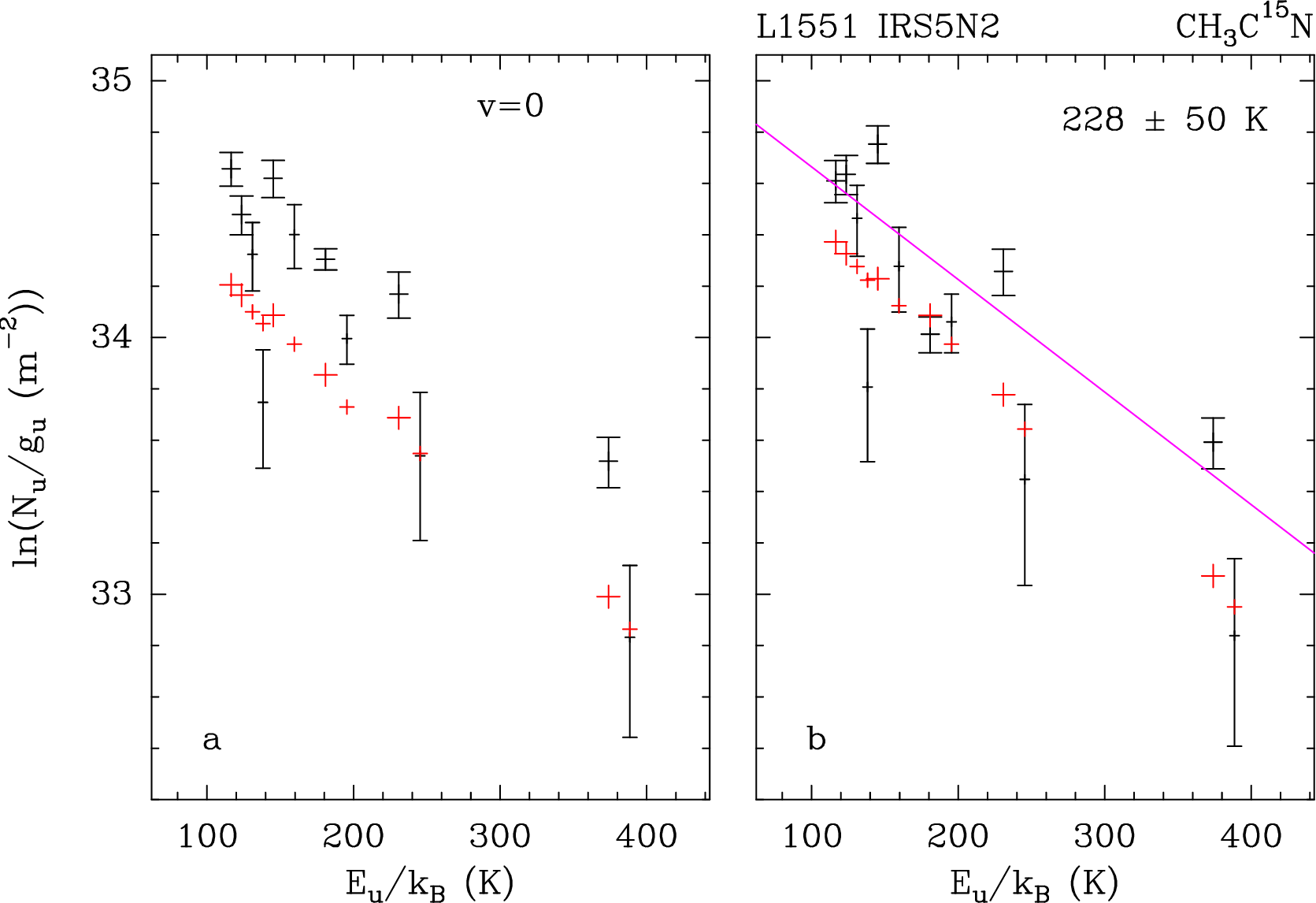}}}
\vspace*{-1ex}
\caption{Same as Fig.~\ref{f:popdiag_ch3cn}, but for CH$_3$C$^{15}$N.}
\label{f:popdiag_ch3cn_15n}
\end{figure}

\begin{figure}[!h]
\centerline{\resizebox{0.80\hsize}{!}{\includegraphics[angle=0]{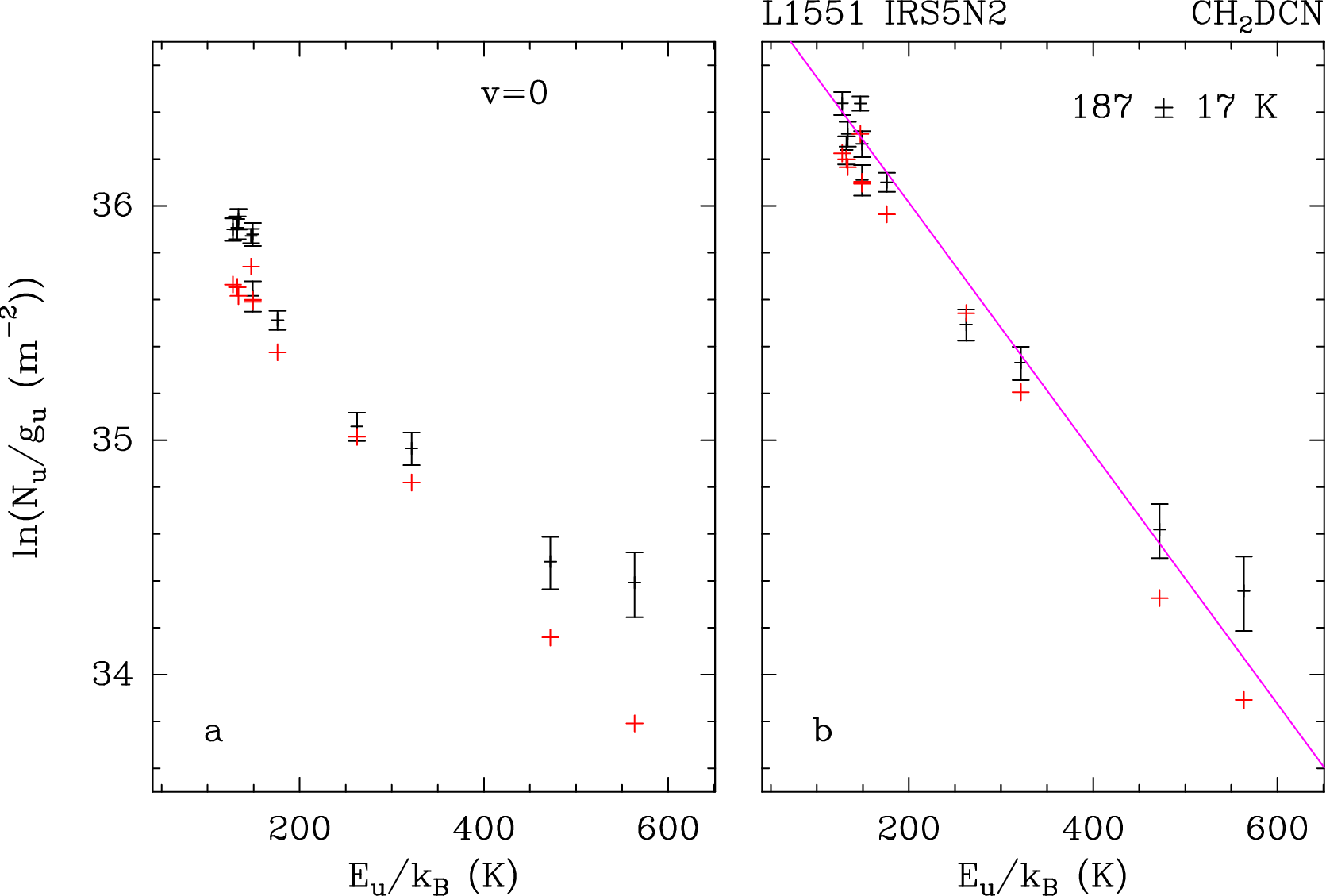}}}
\vspace*{-1ex}
\caption{Same as Fig.~\ref{f:popdiag_ch3cn}, but for CH$_2$DCN.}
\label{f:popdiag_ch2dcn}
\end{figure}

\begin{figure}[!h]
\centerline{\resizebox{0.80\hsize}{!}{\includegraphics[angle=0]{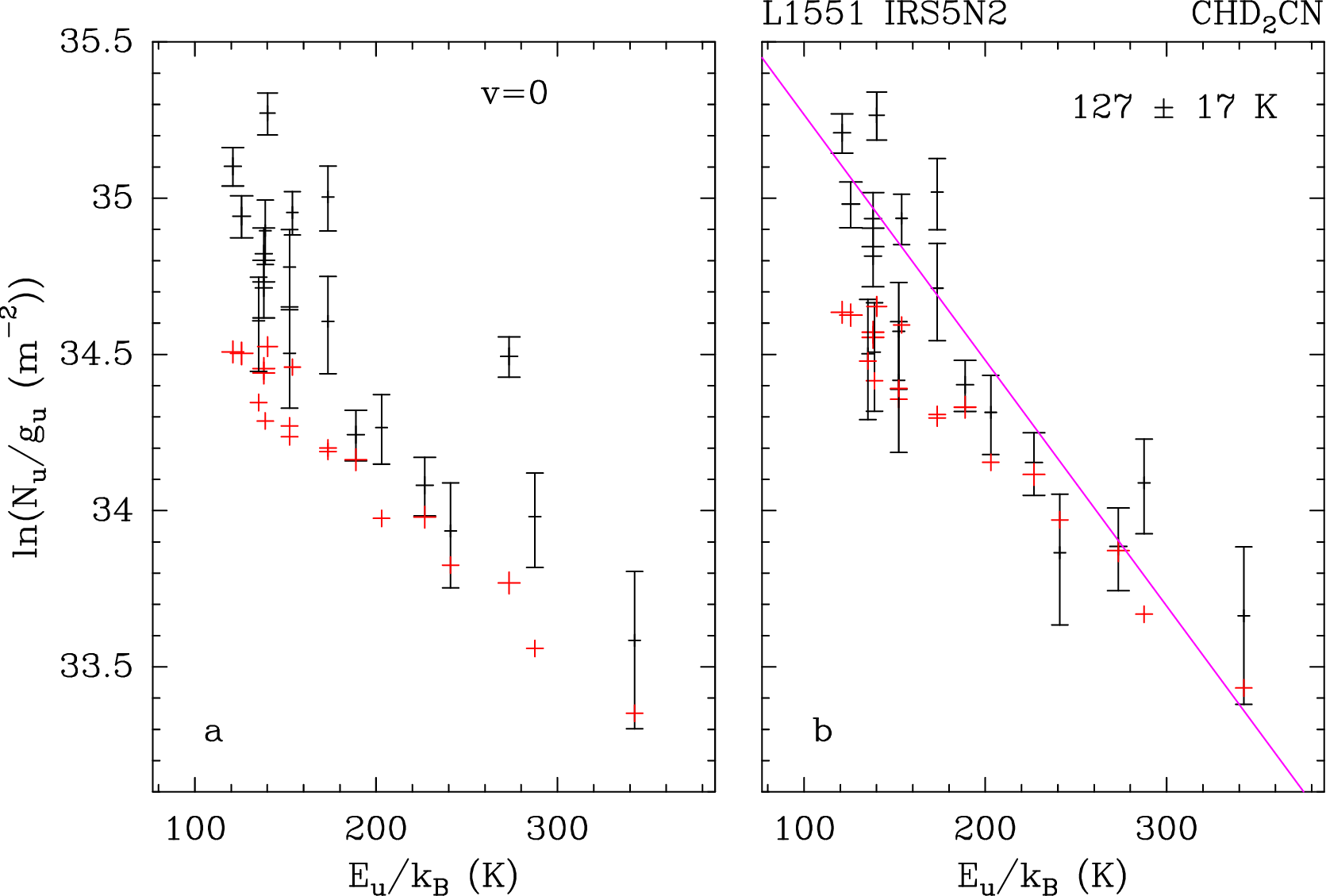}}}
\vspace*{-1ex}
\caption{Same as Fig.~\ref{f:popdiag_ch3cn}, but for CHD$_2$CN.}
\label{f:popdiag_chd2cn}
\end{figure}

\begin{figure}[!h]
\centerline{\resizebox{0.80\hsize}{!}{\includegraphics[angle=0]{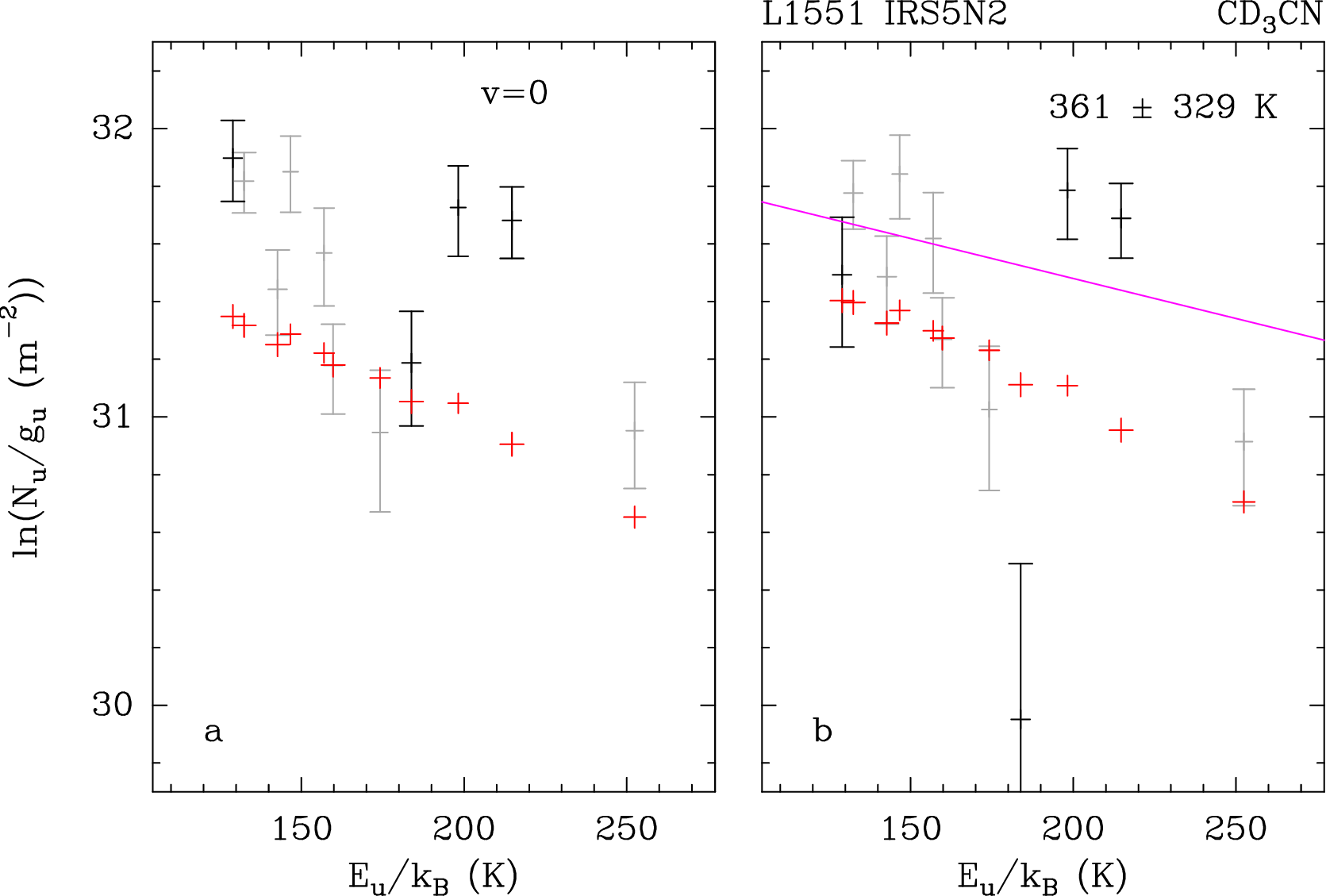}}}
\vspace*{-1ex}
\caption{Same as Fig.~\ref{f:popdiag_ch3cn}, but for CD$_3$CN. The data points corresponding to the lines marked with a star in Fig.~\ref{f:spec_cd3cn_ve0} are highlighted in grey.\newline}
\label{f:popdiag_cd3cn}
\end{figure}

Figures~\ref{f:popdiag_ch3cn}--\ref{f:popdiag_cd3cn} show the population
diagrams of CH$_3$CN,  $^{13}$CH$_3$CN, CH$_3$$^{13}$CN, CH$_3$C$^{15}$N, 
CH$_2$DCN, CHD$_2$CN, and CD$_3$CN, respectively, toward L1551~IRS5N2.

\clearpage
\section{Integrated intensity maps}
\label{a:maps}

\begin{figure}[!h]
\centerline{\resizebox{\hsize}{!}{\includegraphics[angle=0]{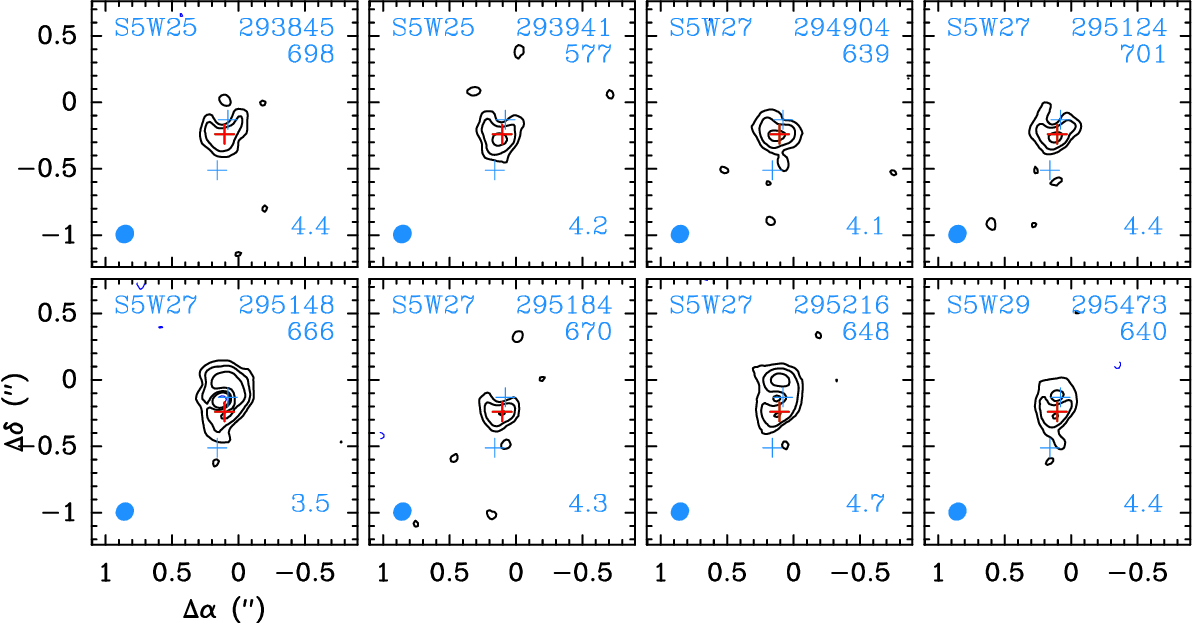}}}
\vspace*{-1ex}
\caption{Integrated intensity maps of CH$_3$CN transitions that are not 
contaminated by emission from other species at the position of the channel 
count peak L1551~IRS5N2 marked with a red cross. The blue crosses indicate the 
continuum peak positions L1551~IRS5N and L1551~IRS5S. In each panel, the setup 
(S) and spectral window (W) numbers are displayed in the top left corner, the 
rest frequency (in MHz) and upper level energy (in K) of the selected 
transition in the top right corner, the noise level 
(in mJy~beam$^{-1}$~km~s$^{-1}$) in the bottom right corner, and the beam in 
the bottom left corner. The contours start at 3 times the noise level 
($\sigma$) and increase by a factor of 2 at each step. Blue dashed contours, 
when present, represent $-3\sigma$.}
\label{f:map_ch3cn}
\end{figure}

\begin{figure}[!h]
\centerline{\resizebox{\hsize}{!}{\includegraphics[angle=0]{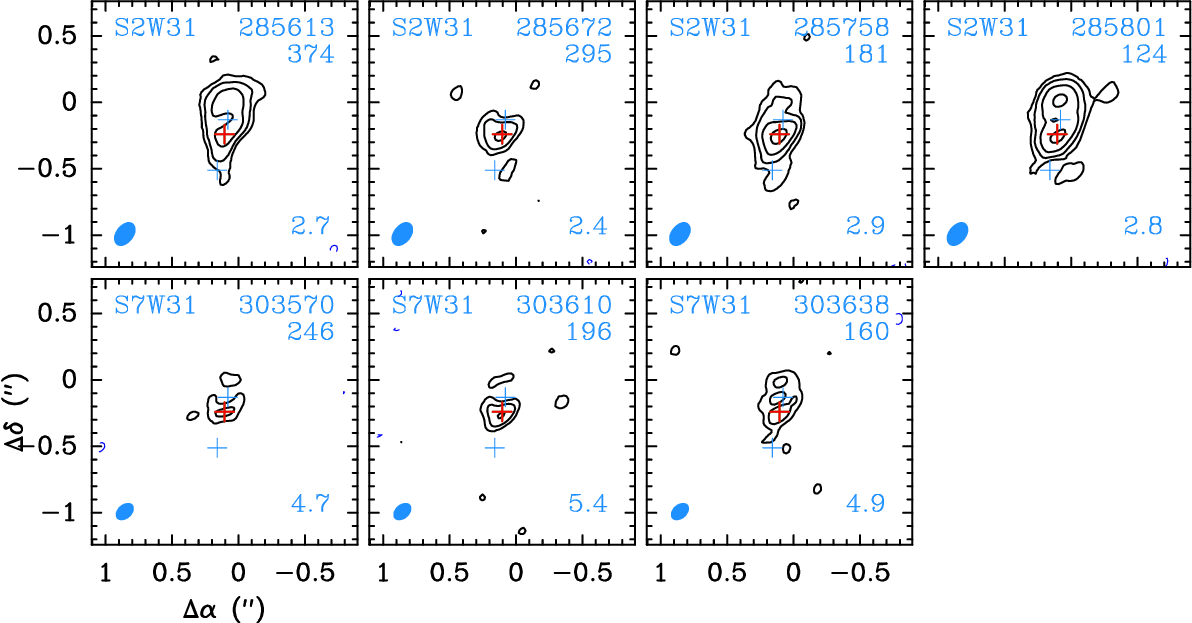}}}
\vspace*{-1ex}
\caption{Same as Fig.~\ref{f:map_ch3cn}, but for $^{13}$CH$_3$CN.}
\label{f:map_ch3cn_13c1}
\end{figure}

\begin{figure}[!h]
\centerline{\resizebox{\hsize}{!}{\includegraphics[angle=0]{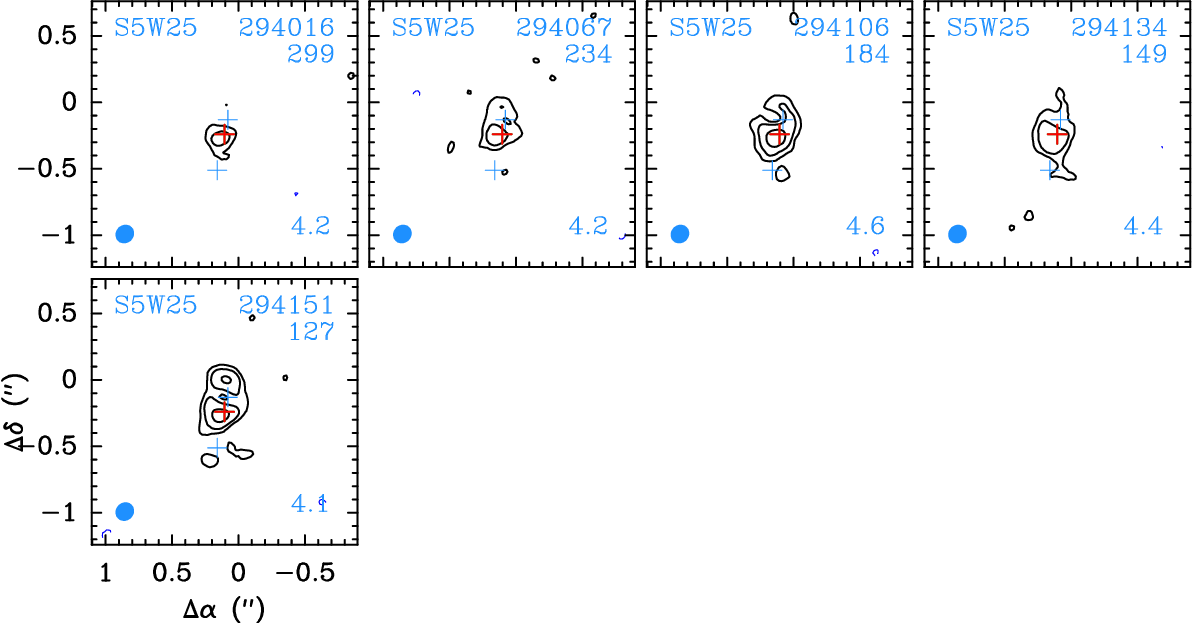}}}
\vspace*{-1ex}
\caption{Same as Fig.~\ref{f:map_ch3cn}, but for CH$_3$$^{13}$CN.}
\label{f:map_ch3cn_13c2}
\end{figure}

\begin{figure}[!h]
\centerline{\resizebox{0.535\hsize}{!}{\includegraphics[angle=0]{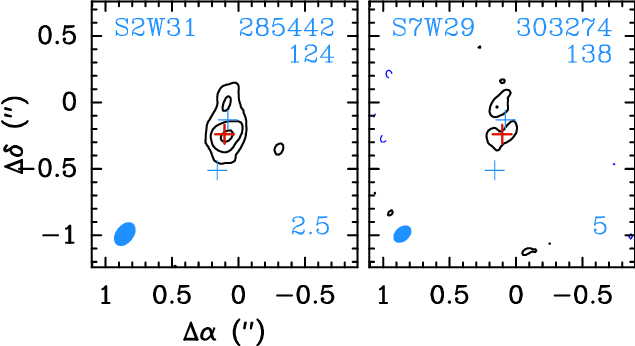}}}
\vspace*{-1ex}
\caption{Same as Fig.~\ref{f:map_ch3cn}, but for CH$_3$C$^{15}$N.}
\label{f:map_ch3cn_15n}
\end{figure}

\begin{figure}[!h]
\centerline{\resizebox{\hsize}{!}{\includegraphics[angle=0]{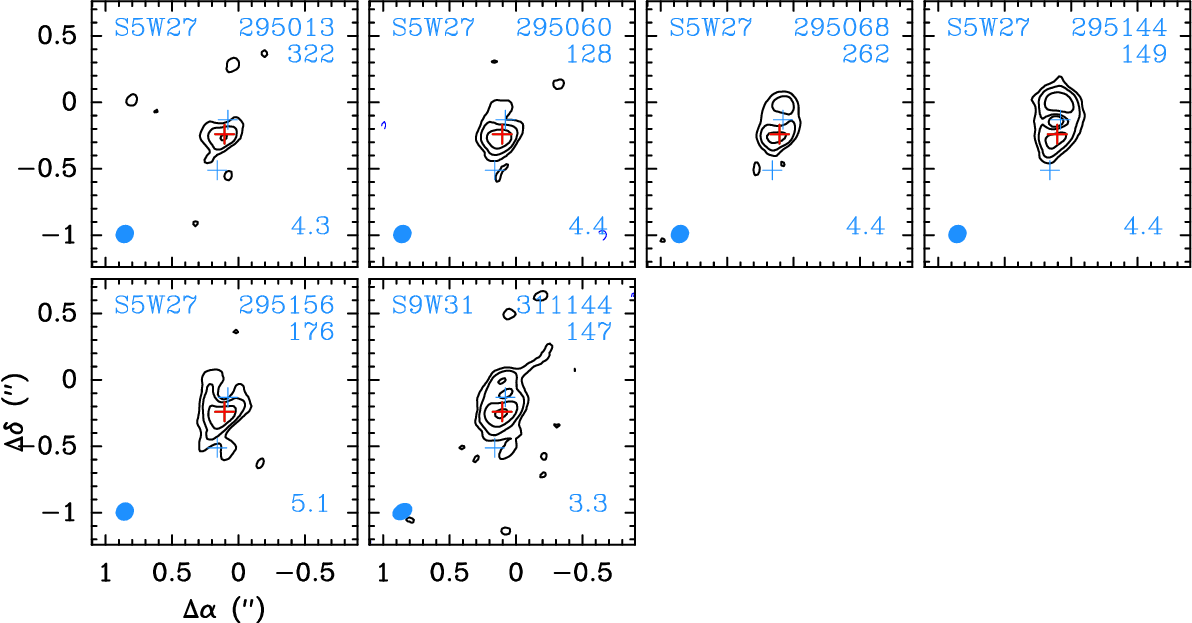}}}
\vspace*{-1ex}
\caption{Same as Fig.~\ref{f:map_ch3cn}, but for CH$_2$DCN.}
\label{f:map_ch2dcn}
\end{figure}

\begin{figure}[!h]
\centerline{\resizebox{0.765\hsize}{!}{\includegraphics[angle=0]{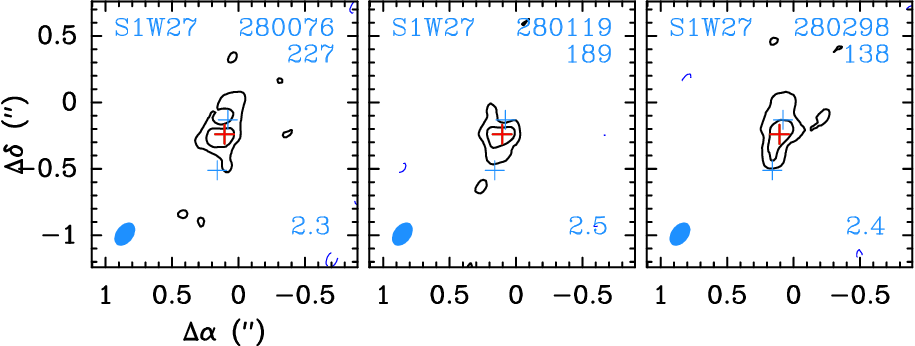}}}
\vspace*{-1ex}
\caption{Same as Fig.~\ref{f:map_ch3cn}, but for CHD$_2$CN.}
\label{f:map_chd2cn}
\end{figure}

\begin{figure}[!h]
\centerline{\resizebox{\hsize}{!}{\includegraphics[angle=0]{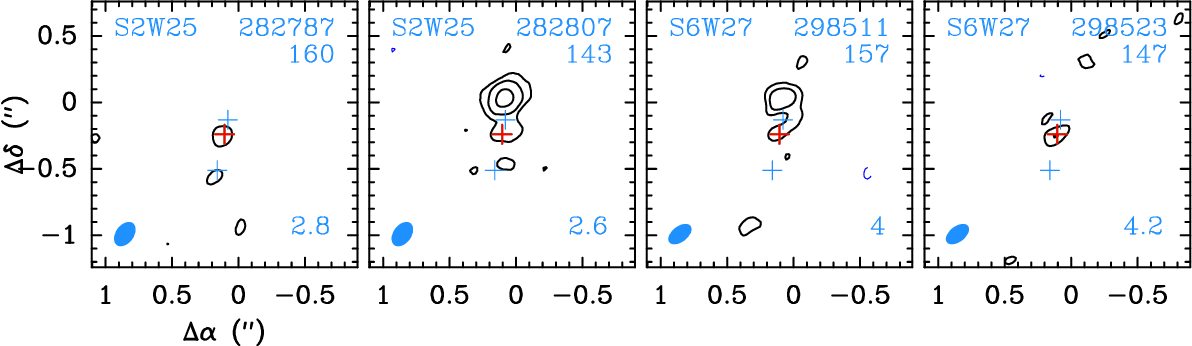}}}
\vspace*{-1ex}
\caption{Same as Fig.~\ref{f:map_ch3cn}, but for CD$_3$CN.}
\label{f:map_cd3cn}
\end{figure}

Figures~\ref{f:map_ch3cn}--\ref{f:map_cd3cn} show integrated intensity maps
of CH$_3$CN,  $^{13}$CH$_3$CN, CH$_3$$^{13}$CN, CH$_3$C$^{15}$N, 
CH$_2$DCN, CHD$_2$CN, and CD$_3$CN, respectively, toward L1551~IRS5.
Because of the large velocity gradient across the source, any emission that
shows up to the north of L1551~IRS5N must trace other molecules (see 
Sect.~\ref{ss:param}).

\end{appendix}

\end{document}

%% file: tab_l1551irs5_p2_cd3cn_popfit.tex
\begin{table}
 \begin{center}
 \caption{
 Rotational temperature of methyl cyanide isotopologs derived from their population diagrams toward L1551~IRS5N2.
}
 \label{t:popfit}
 \vspace*{0.0ex}
 \begin{tabular}{lllll}
 \hline\hline
 \multicolumn{1}{c}{Molecule} & \multicolumn{1}{c}{\hspace*{-1.5ex} States\tablefootmark{(a)}} & \multicolumn{3}{c}{$T_{\rm fit}$ \small (K)\tablefootmark{(b)}} \\ 
 \cline{3-5}
  & & \multicolumn{1}{c}{\hspace*{-1.5ex} Best\tablefootmark{(c)}} & \multicolumn{1}{c}{\hspace*{-1.ex} Alt. 1\tablefootmark{(d)}} & \multicolumn{1}{c}{\hspace*{-1.ex} Alt. 2\tablefootmark{(e)}} \\ 
 \hline
CH$_3$CN & \hspace*{-1.5ex} $\varv=0$, $\varv_8=1$ & \hspace*{-1.ex}   215 (19) & \hspace*{-1.ex}   254 (30) & \hspace*{-1.ex}   189 (16) \\ 
\hline 
$^{13}$CH$_3$CN & \hspace*{-1.5ex} $\varv=0$, $\varv_8=1$ & \hspace*{-1.ex}   183 (21) & \hspace*{-1.ex}   199 (28) & \hspace*{-1.ex}   178 (20) \\ 
CH$_3$$^{13}$CN & \hspace*{-1.5ex} $\varv=0$ & \hspace*{-1.ex}   179 (31) & \hspace*{-1.ex}   204 (42) & \hspace*{-1.ex}   174 (29) \\ 
CH$_3$C$^{15}$N & \hspace*{-1.5ex} $\varv=0$ & \hspace*{-1.ex}   228 (50) & \hspace*{-1.ex}   237 (58) & \hspace*{-1.ex}   223 (48) \\ 
\hline 
CH$_2$DCN & \hspace*{-1.5ex} $\varv=0$ & \hspace*{-1.ex}   187 (17) & \hspace*{-1.ex}   208 (22) & \hspace*{-1.ex}   181 (18) \\ 
CHD$_2$CN & \hspace*{-1.5ex} $\varv=0$ & \hspace*{-1.ex}   127 (17) & \hspace*{-1.ex}   127 (17) & \hspace*{-1.ex}   124 (16) \\ 
CD$_3$CN & \hspace*{-1.5ex} $\varv=0$ & \hspace*{-1.ex}   360 (330) & \hspace*{-1.ex}   380 (370) & \hspace*{-1.ex}   350 (320) \\ 
\hline 
 \end{tabular}
 \end{center}
 \vspace*{-2.5ex}
 \tablefoot{
 \tablefoottext{a}{Vibrational states that were taken into account to fit the population diagram.}
 \tablefoottext{b}{The standard deviation of the fit is given in parentheses. As explained in Sect.~3 of \citet{Belloche16} and in Sect.~4.4 of \citet{Belloche19}, this uncertainty is purely statistical and should be viewed with caution. It may underestimate the actual uncertainty.}
 \tablefoottext{c}{With corrections for opacity and contamination based on the best-fit model.}
 \tablefoottext{d}{With corrections for opacity and contamination based on the alternative model with an emission size of 0.15$\arcsec$ and a temperature of 200~K.}
 \tablefoottext{e}{With corrections for opacity and contamination based on the alternative model with a size of 0.15$\arcsec$ and a temperature of 150~K.}
 }
 \end{table}

%% file: tab_l1551irs5_p2_cd3cn_weedsmodel.tex
\begin{table*}[!ht]
 \begin{center}
 \caption{
 Parameters of our best-fit and alternative LTE models of methyl cyanide and its isotopologs toward L1551~IRS5N2.
}
 \label{t:coldens}
 \vspace*{-1.2ex}
 \begin{tabular}{lcrcccccc}
 \hline\hline
 \multicolumn{1}{c}{Molecule, vib. state} & \multicolumn{1}{c}{Status\tablefootmark{a}} & \multicolumn{1}{c}{$N_{\rm det}$\tablefootmark{b}} & \multicolumn{1}{c}{$N$\tablefootmark{c}} & \multicolumn{1}{c}{$C_{\mathrm{vib}}$\tablefootmark{e}} & \multicolumn{1}{c}{$\frac{N_{12}}{N}$\tablefootmark{f}} & \multicolumn{1}{c}{$\frac{N}{N_{12}}$\tablefootmark{f}} & \multicolumn{1}{c}{$\frac{68 \times N_{13}}{N}$\tablefootmark{g}} & \multicolumn{1}{c}{$\frac{N}{68 \times N_{13}}$\tablefootmark{g}} \\ 
  & & & \multicolumn{1}{c}{\small (cm$^{-2}$)} & & & \multicolumn{1}{c}{\small (\%)} & & \multicolumn{1}{c}{\small (\%)} \\ 
 \hline
\smallskip \\[-2.ex] \multicolumn{8}{c}{Best-fit model: size  0.09$\arcsec$, temperature  200~K} \\ \smallskip \\[-2.ex]  CH$_3$CN, $\varv=0$ & d & 11 &  5.0 (16)\tablefootmark{d} & 1.00 &       1 &  \ldots & \ldots & \ldots \\ 
 \hspace*{7.7ex} $\varv_8=1$ & d & 11 &  5.0 (16)\tablefootmark{d} & 1.00 &       1 &  \ldots & \ldots & \ldots \\ 
 \hspace*{7.7ex} $\varv_8=2$ & n & 0 &  5.0 (16)\tablefootmark{d} & 1.00 &       1 &  \ldots & \ldots & \ldots \\ 
\hline 
 $^{13}$CH$_3$CN, $\varv=0$ & d & 13 &  4.1 (15) & 1.18 &      12 &  \ldots &      68 &  \ldots \\ 
 \hspace*{9.4ex} $\varv_8=1$ & n & 0 &  4.1 (15) & 1.18 &      12 &  \ldots &      68 &  \ldots \\ 
 CH$_3$$^{13}$CN, $\varv=0$ & d & 7 &  4.1 (15) & 1.18 &      12 &  \ldots &      68 &  \ldots \\ 
 \hspace*{9.4ex} $\varv_8=1$ & n & 0 &  4.1 (15) & 1.18 &      12 &  \ldots &      68 &  \ldots \\ 
 CH$_3$C$^{15}$N, $\varv=0$ & d & 10 &  1.5 (15) & 1.18 &      33 &  \ldots &     180 &  \ldots \\ 
 \hspace*{9.4ex} $\varv_8=1$ & n & 0 &  1.5 (15) & 1.18 &      33 &  \ldots &     180 &  \ldots \\ 
 CH$_2$DCN, $\varv=0$ & d & 13 &  9.4 (15) & 1.18 &     5.3 &      19 &      30 &     3.4 \\ 
 CHD$_2$CN, $\varv=0$ & d & 15 &  2.4 (15) & 1.22 &      21 &     4.9 &     110 &    0.87 \\ 
 CD$_3$CN, $\varv=0$ & d & 7 &  9.7 (14) & 1.22 &      51 &     1.9 &     290 &    0.35 \\ 
\hline 
\smallskip \\[-2.ex] \multicolumn{8}{c}{Alternative model: size  0.15$\arcsec$, temperature  200~K} \\ \smallskip \\[-2.ex]  CH$_3$CN, $\varv=0$ & d & 10 &  2.0 (16)\tablefootmark{d} & 1.00 &       1 &  \ldots & \ldots & \ldots \\ 
 \hspace*{7.7ex} $\varv_8=1$ & d & 8 &  2.0 (16)\tablefootmark{d} & 1.00 &       1 &  \ldots & \ldots & \ldots \\ 
 \hspace*{7.7ex} $\varv_8=2$ & n & 0 &  2.0 (16)\tablefootmark{d} & 1.00 &       1 &  \ldots & \ldots & \ldots \\ 
\hline 
 $^{13}$CH$_3$CN, $\varv=0$ & d & 13 &  1.6 (15) & 1.18 &      12 &  \ldots &      68 &  \ldots \\ 
 \hspace*{9.4ex} $\varv_8=1$ & n & 0 &  1.6 (15) & 1.18 &      12 &  \ldots &      68 &  \ldots \\ 
 CH$_3$$^{13}$CN, $\varv=0$ & d & 7 &  1.6 (15) & 1.18 &      12 &  \ldots &      68 &  \ldots \\ 
 \hspace*{9.4ex} $\varv_8=1$ & n & 0 &  1.6 (15) & 1.18 &      12 &  \ldots &      68 &  \ldots \\ 
 CH$_3$C$^{15}$N, $\varv=0$ & d & 10 &  7.6 (14) & 1.18 &      26 &  \ldots &     150 &  \ldots \\ 
 \hspace*{9.4ex} $\varv_8=1$ & n & 0 &  7.6 (14) & 1.18 &      26 &  \ldots &     150 &  \ldots \\ 
 CH$_2$DCN, $\varv=0$ & d & 12 &  4.1 (15) & 1.18 &     4.9 &      21 &      27 &     3.7 \\ 
 CHD$_2$CN, $\varv=0$ & d & 15 &  1.2 (15) & 1.22 &      16 &     6.1 &      92 &     1.1 \\ 
 CD$_3$CN, $\varv=0$ & d & 7 &  4.9 (14) & 1.22 &      41 &     2.4 &     230 &    0.44 \\ 
\hline 
\smallskip \\[-2.ex] \multicolumn{8}{c}{Alternative model: size  0.15$\arcsec$, temperature  150~K} \\ \smallskip \\[-2.ex]  CH$_3$CN, $\varv=0$ & d & 10 &  6.5 (16)\tablefootmark{d} & 1.00 &       1 &  \ldots & \ldots & \ldots \\ 
 \hspace*{7.7ex} $\varv_8=1$ & d & 7 &  6.5 (16)\tablefootmark{d} & 1.00 &       1 &  \ldots & \ldots & \ldots \\ 
 \hspace*{7.7ex} $\varv_8=2$ & n & 0 &  6.5 (16)\tablefootmark{d} & 1.00 &       1 &  \ldots & \ldots & \ldots \\ 
\hline 
 $^{13}$CH$_3$CN, $\varv=0$ & d & 13 &  2.0 (15) & 1.07 &      32 &  \ldots &      68 &  \ldots \\ 
 \hspace*{9.4ex} $\varv_8=1$ & n & 0 &  2.0 (15) & 1.07 &      32 &  \ldots &      68 &  \ldots \\ 
 CH$_3$$^{13}$CN, $\varv=0$ & d & 7 &  2.0 (15) & 1.07 &      32 &  \ldots &      68 &  \ldots \\ 
 \hspace*{9.4ex} $\varv_8=1$ & n & 0 &  2.0 (15) & 1.07 &      32 &  \ldots &      68 &  \ldots \\ 
 CH$_3$C$^{15}$N, $\varv=0$ & d & 9 &  8.0 (14) & 1.07 &      81 &  \ldots &     170 &  \ldots \\ 
 \hspace*{9.4ex} $\varv_8=1$ & n & 0 &  8.0 (14) & 1.07 &      81 &  \ldots &     170 &  \ldots \\ 
 CH$_2$DCN, $\varv=0$ & d & 10 &  5.0 (15) & 1.07 &      13 &     7.7 &      27 &     3.6 \\ 
 CHD$_2$CN, $\varv=0$ & d & 13 &  1.3 (15) & 1.09 &      50 &     2.0 &     110 &    0.95 \\ 
 CD$_3$CN, $\varv=0$ & d & 7 &  4.9 (14) & 1.09 &     130 &    0.75 &     280 &    0.36 \\ 
\hline 
 \end{tabular}
 \end{center}
 \vspace*{-2.5ex}
 \tablefoot{
 All entries were modeled assuming a line width (FWHM) of 1.5~km~s$^{-1}$ and a velocity of 5.8~km~s$^{-1}$.
 \tablefoottext{a}{d: detection, n: nondetection.}
 \tablefoottext{b}{Number of detected lines \citep[conservative estimate, see Sect.~3 of][ and Sect.~3.3 in this work]{Belloche16}. One line of a given species may mean a group of transitions of that species that are blended together.}
 \tablefoottext{c}{Total column density of the molecule. $x$ ($y$) means $x \times 10^y$. An identical value for the various vibrational states means that LTE describes well their relative populations, that is that the vibrational temperature is equal to the rotational temperature (see Sect. 2.3 of \citet{Belloche25}).}
 \tablefoottext{d}{This value likely underestimates the column density of the main isotopolog (see Sect.~\ref{ss:isoratios}).}
 \tablefoottext{e}{Correction factor that was applied to the column density to account for the contribution of vibrationally excited states, in the cases where this contribution was not included in the partition function of the spectroscopic predictions.}
 \tablefoottext{f}{Column density ratio, with $N_{12}$ the column density of CH$_3$CN, which is likely underestimated (see Sect.~\ref{ss:isoratios}).}
 \tablefoottext{g}{Column density ratio, with $N_{13}$ the column density of $^{13}$CH$_3$CN and 68 the assumed $\frac{^{12}{\rm C}}{^{13}{\rm C}}$ isotopic ratio.}
 }
 \end{table*}

%% file: tab_l1551irs5_p2_cd3cn_deuteration.tex
\begin{table}[!h]
 \begin{center}
 \caption{
 Deuteration level of CH$_3$CN toward L1551~IRS5N2.
}
 \label{t:deuteration}
 \vspace*{-1.2ex}
 \begin{tabular}{lcc}
 \hline\hline
 \multicolumn{1}{c}{Molecules} & \multicolumn{1}{c}{$R$ (\%)\tablefootmark{a}} & \multicolumn{1}{c}{D/H (\%)\tablefootmark{b}} \\ 
 \hline
 CH$_2$DCN/CH$_3$CN\tablefootmark{c} &     3.4 $\pm$     0.8 &    1.1 $\pm$     0.3 \\ 
 CHD$_2$CN/CH$_3$CN\tablefootmark{c} &    0.87 $\pm$    0.22 &    5.4 $\pm$     0.7 \\ 
 CD$_3$CN/CH$_3$CN\tablefootmark{c} &    0.35 $\pm$    0.10 &   15.2 $\pm$     1.5 \\ 
 CHD$_2$CN/CH$_2$DCN &      26 $\pm$       4 &     26 $\pm$       4 \\ 
 CD$_3$CN/CHD$_2$CN &      40 $\pm$      10 &    120 $\pm$      30 \\ 
 \hline
 \end{tabular}
 \end{center}
 \vspace*{-2.5ex}
 \tablefoot{
 \tablefoottext{a}{Column density ratio, computed with the column densities of the best-fit model reported in Table~\ref{t:coldens}.}
 \tablefoottext{b}{D/H ratio assuming a purely statistical distribution.}
 \tablefoottext{c}{Ratio computed using 68$\times$$N$($^{13}$CH$_3$CN) as a proxy for $N$(CH$_3$CN).}
 }
 \end{table}